\documentclass[a4paper,10pt,twoside]{article}

\usepackage{geometry}	
\usepackage{xcolor}
\definecolor{orange}{RGB}{255, 222, 173}
\definecolor{uglyblue}{RGB}{95,158,160}
\definecolor{newblue}{RGB}{128,0,0}
\definecolor{mygray}{RGB}{220,220,220}
\definecolor{mywhite}{RGB}{255,250,240}
\definecolor{DarkRed}{rgb}{0.6,0,0}
\definecolor{DarkGreen}{rgb}{0,0.6,0}
\definecolor{DarkBlue}{rgb}{0,0,0.6}

\usepackage{setspace}

\everydisplay{\def\arraystretch{0.8}}

\usepackage{tocloft}

\renewcommand\thesection{\arabic{section}}

\makeatletter
\def\hlinewd#1{%
	\noalign{\ifnum0=`}\fi\hrule \@height #1 %
	\futurelet\reserved@a\@xhline}
\makeatother

\usepackage{verbatim}
 
\usepackage[colorlinks=true,linktocpage=true,linkcolor=DarkBlue,citecolor=DarkGreen,urlcolor=DarkGreen]{hyperref}

\usepackage{array}

\usepackage[latin9,utf8]{inputenc} % latin9 input encoding
\usepackage[T1]{fontenc}
\usepackage{bm}
\usepackage{cancel}
\usepackage{float,hypcap}
\usepackage{changepage,titlesec}
\usepackage{soul}
\usepackage{bbold}
\usepackage{slashed}
\usepackage{tabularx}
\usepackage{amsmath,amsfonts,amssymb}
\usepackage{amsmath,bbm,latexsym,amssymb}
\usepackage{braket}
\usepackage{blkarray}
\usepackage{enumerate}
\usepackage{booktabs,hyperref}
\usepackage{bigints}
\usepackage{cite}
\usepackage{mflogo}
\usepackage{scrextend}
\usepackage{longtable}
\usepackage{enumitem}
\usepackage{pifont}
\usepackage{cases}
\usepackage{colortbl}
\usepackage[export]{adjustbox}
\usepackage{arydshln}
\usepackage{colortbl}
\usepackage{graphics}

\usepackage{multirow}

\usepackage[flushleft]{threeparttable}

\usepackage{verbatim}

\usepackage{fancyvrb}
\usepackage{xcolor}

\usepackage{graphicx}
\usepackage{caption}
\usepackage{subfig}
\usepackage{feynmp-auto}

\usepackage[most]{tcolorbox}

\let\oldcdot\cdot 
\usepackage{breqn} 
\let\cdot\oldcdot 
\usepackage{authblk}

\allowdisplaybreaks

\def\be{\begin{equation}}
\def\ee{\end{equation}}
\def\ba{\begin{alignedat}}
\def\ea{\end{alignedat}}
\def\bea{\begin{eqnarray}}
\def\eea{\end{eqnarray}}
\newcommand{\bs}{\begin{subequations}}
\newcommand{\es}{\end{subequations}}

\def\vs{\vspace}

\def\no{\nonumber\\}
\def\fn{\footnote}

\newcommand{\newc}{\newcommand}
\newc{\ol}{\overline}
\newc{\wt}{\widetilde}

\newc{\m}{\mathcal}

\newcommand{\beq}{\begin{eqnarray}}
\newcommand{\eeq}{\end{eqnarray}}
\newcommand{\bpmatrix}{\begin{pmatrix}}
\newcommand{\epmatrix}{\end{pmatrix}}

\renewcommand{\ol}{\text{1l}}

\renewcommand{\eqref}[1]{Eq.~(\ref{#1})}

\newcommand{\bc}{\begin{center}}
\newcommand{\ec}{\end{center}}
\newcommand{\lsim}{\raisebox{-0.13cm}{~\shortstack{$<$ \\[-0.07cm]
      $\sim$}}~}

\def\m#1{m_{#1}}

\def\ltap{\;\centeron{\raise.35ex\hbox{$<$}}{\lower.65ex\hbox{$\sim$}}\;}
\def\gtap{\;\centeron{\raise.35ex\hbox{$>$}}{\lower.65ex\hbox{$\sim$}}\;}

\usepackage{stackengine}
\stackMath

\usepackage{lscape}

\usepackage{imakeidx}
\makeindex
\usepackage{makecell}
\usepackage{fontawesome}

\let\OLDthebibliography\thebibliography
\renewcommand\thebibliography[1]{
	\OLDthebibliography{#1}
	\setlength{\parskip}{3.0pt plus 2.5pt minus 1.0pt}
	\setlength{\itemsep}{1.0pt plus 2.5pt minus 1.0pt}
}

\definecolor{neworange}{rgb}{1,0.5,0}

\definecolor{mygreen}{RGB}{0,128,0}

\definecolor{raq}{RGB}{255,164,0}

\newcommand{\cba}{c_{\beta \scalebox{0.5}[0.85]{$-$} \alpha}}
\newcommand{\sba}{s_{\beta \scalebox{0.5}[0.85]{$-$} \alpha}}
\newcommand{\lrDmu}{\overset{\scalebox{1.1}[0.75]{$\leftrightarrow$}}{D}_{\!\mu}}
\newcommand{\lrDmuI}{\overset{\scalebox{1.1}[0.75]{$\leftrightarrow$}}{D}_{\!\mu}{}^{\!\!\!I}}

\newcommand{\mym}{\hspace{-2mm}}

\newcommand{\nnl}{\eta_e}
\newcommand{\nnd}{\eta_d}
\newcommand{\nnu}{\eta_u}
\begin{document}

% ====================================== Page Numbering
\pagestyle{plain}
\hypersetup{pageanchor=false}
% ====================================== Page Numbering II
\hypersetup{pageanchor=true}
\pagenumbering{roman}
\setcounter{page}{1}
% ====================================== Page Numbering III
\pagenumbering{arabic}
\setcounter{page}{1}

% ====================================== Title part
\makeatletter   
\renewcommand\maketitle{
{\raggedright % Note the extra {
\begin{center}
{\large \bfseries \@title }

\vspace{1mm}

{\@author}

\vspace{1.2mm}

%(\@date)
\end{center}}} % Note the extra }
\makeatother

\title{Relevance of one-loop SMEFT matching in the 2HDM}
\renewcommand*{\thefootnote}{\fnsymbol{footnote}}
\author[1]{Supratim Das Bakshi\footnote{\href{mailto:sdb@ugr.es}{sdb@ugr.es}}}
\author[2]{Sally Dawson\footnote{\href{mailto:dawson@bnl.gov}{dawson@bnl.gov}}}
\author[2]{Duarte Fontes\footnote{\href{mailto:dfontes@bnl.gov}{dfontes@bnl.gov}}}
\author[3]{Samuel Homiller\footnote{\href{mailto:shomiller@g.harvard.edu}{shomiller@g.harvard.edu}\\}}
\affil[1]{\textit{CAFPE and Departamento de Física Teórica y del Cosmos, Universidad de Granada, \linebreak Campus de Fuentenueva, E–18071 Granada, Spain}}
\affil[2]{\textit{Department of Physics, Brookhaven National Laboratory, Upton, New York, 11973, U.S.A.}}
\affil[3]{\textit{Physics Department, Harvard University, Cambridge, MA, 02138, U.S.A.}}

\date{\today}

\maketitle

\renewcommand*{\thefootnote}{\arabic{footnote}}
\setcounter{footnote}{0}

\vs{0.5mm}
\begin{addmargin}[12mm]{12mm}
\small
The Two-Higgs Doublet Model (2HDM) is a well understood alternative to the Standard Model of particle physics. If the new particles included in the 2HDM are at an energy scale much greater than the weak scale, the theory can be matched to the Standard Model Effective Field Theory (SMEFT). We compute for the first time the complete one-loop matching at dimension-6. We compare its numerical impact with that of tree-level matching at dimension-8 by performing a global fit to single Higgs and precision electroweak measurements, and we emphasize the importance of comparing one-loop SMEFT results with corresponding one-loop results in the full 2HDM model.
In the SMEFT, we consider the relative importance of both one-loop matching and the inclusion of renormalization group evolution.
Our results demonstrate the necessity of studying the impact of various expansions to quantify the uncertainties of the SMEFT matching. 
\end{addmargin}

\normalsize

%\vspace{15mm}

\section{Introduction}
\label{sec:Introduction}

Since the observation at the LHC of a light Higgs boson with properties approximately those predicted by the Standard Model (SM), much of the focus has turned to searches for new physics at higher energy scales. These studies follow a two pronged approach: one either searches directly for heavy new particles that are typically predicted in extensions of the SM, or uses an effective field theory (EFT) framework to look for weak scale manifestations of the new physics. Here we follow the second alternative, resorting to the Standard Model Effective Field Theory (SMEFT) (for reviews, see Refs.~\cite{Brivio:2017vri,Isidori:2023pyp}). The SMEFT contains an infinite tower of higher-dimensional $\textrm{SU}(3)_c\times \textrm{SU}(2)_L \times \textrm{U}(1)_Y$ invariant operators,%
\fn{We are assuming lepton and baryon number conservation, which means we only keep operators of even mass dimension.}
\begin{equation}
\mathcal{L}_{\rm{SMEFT}} = \mathcal{L}_{\rm{SM}} + \sum_{i}\frac{C_i O_i}{\Lambda^{2}} + \sum_{j} \frac{C_j O_j}{\Lambda^{4}} + ... \, \, ,
\end{equation}
where $\mathcal{L}_{\rm{SM}}$ is the renormalizable SM Lagrangian, $\Lambda$ is the scale of some conjectured ultraviolet (UV) complete model and the $C_i$ are usually known as Wilson coefficients (WCs). In any given model of UV physics, the WCs can be calculated in terms of the model parameters.

The goal of the SMEFT program is to experimentally observe a pattern of non-zero WCs, and thus infer features of the physics at the high scale. It is then of paramount importance to investigate how well the WCs can replicate the UV physics at low energy scales. A usual strategy is to flip the problem upside down, by starting from a particular UV model, and by performing the matching between that model and the SMEFT \cite{Henning:2014wua, Gorbahn:2015gxa, Jiang:2018pbd, Haisch:2020ahr, Dawson:2020oco, DasBakshi:2020ejz}. 
Although a tree-level matching with dimension-6 operators is the obvious first step, it is clear that there are scenarios where this is insufficient~\cite{Contino:2016jqw, Belusca-Maito:2016cay, Ellis:2019zex, Li:2020gnx, Murphy:2020rsh, Corbett:2021eux, Degrande:2023iob, Corbett:2023qtg}. Improvements have been made either in the direction of increasing the dimension \cite{Dawson:2021xei, Dawson:2022cmu, Banerjee:2022thk, Ellis:2023zim}, or in the direction of one-loop matching \cite{Jiang:2018pbd, Haisch:2020ahr, Anisha:2020ggj, Du:2022vso, Liao:2022cwh, Li:2023ohq}. 
The complexity of the matching procedure and interpretation of data  increases with each of these improvements.
A relevant question is then: which of these directions is necessary to quickly approach the UV model using the SMEFT approach?

We address this question by considering the Two-Higgs Doublet Model (2HDM)~\cite{Lee:1973iz} with a softly broken $Z_2$ symmetry. This model is an excellent test case, since the tree-level matching to the SMEFT up to dimension-8 is known~\cite{Dawson:2022cmu}, and the phenomenological consequences of the full 2HDM model are well studied~\cite{Branco:2011iw, Gunion:2002zf}. The matching procedure of a UV model like the 2HDM to the SMEFT mainly follows two approaches: diagrammatic and functional methods.%
\fn{In recent studies, the on-shell amplitude methods are also being developed, see Refs.~\cite{DeAngelis:2023bmd,Li:2023edf,Aebischer:2023nnv}.}
The diagrammatic approach involves the computation of amplitudes in the UV and in the EFT, and solving for WCs in terms of the new physics couplings. The functional methods proceed with the computation of effective action formulae, defined by the covariant derivatives and the new physics interactions. Both methods involve lengthy and cumbersome mathematics, where automatization can help immensely. Currently, there are four packages that facilitate automated SMEFT matching: \texttt{MatchingTools} \cite{Criado:2017khh}, \texttt{CoDEx} \cite{DasBakshi:2018vni}, \texttt{Matchmakereft} \cite{Carmona:2021xtq} and \texttt{Matchete} \cite{Fuentes-Martin:2022jrf}. 

In this work, we perform the complete one-loop matching of the 2HDM to the SMEFT with dimension-6 operators, using the packages \texttt{Matchmakereft} and \texttt{Matchete}. We create an implementation of the 2HDM for these software packages, and check that their results agree. The results of these two codes have been checked to agree in many other scenarios, including Higgs singlet model \cite{Haisch:2020ahr}, vector-like lepton extension of the SM \cite{Fuentes-Martin:2022jrf}, Type-III seesaw model \cite{Li:2023ohq} and a lepto-quark model \cite{Gherardi:2020det}.  
We also use \texttt{Matchete} to calculate the tree-level dimension-8 WCs, by exploiting the capacity of this software to create and solve effective action terms~\cite{Fuentes-Martin:2020udw} %please add \usepackage[normalem]{ulem} for \sout command
at arbitrary mass-dimension, without additional inputs from the users.
The WCs which are generated at tree-level (both dimension-6 and dimension-8) and which are relevant for this study fully agree with those of Ref.~\cite{Dawson:2022cmu}. 

Here, we are interested in using the 2HDM to test the quality of SMEFT matchings. Specifially, we  compare the numerical importance of one-loop matching with dimension-6 operators, on the one hand, and tree-level matching with dimension-8 operators, on the other. This we do by examining both precision electroweak and single Higgs measurements. It is of particular importance to compare the SMEFT one-loop matched results to predictions for observables in the full 2HDM that are also computed to one-loop~\cite{Chen:2019pkq,Chen:2018shg}.
A similar study performed for a Higgs singlet model found that the effects of the one-loop dimension-6 matching in that case were small~\cite{Dawson:2021jcl}.

The paper is organized as follows. In Section~\ref{sec:twhdm}, we briefly summarize the 2HDM in order to set notation, and we discuss the constraints on the parameters of the model.
Section~\ref{sec:matching} is devoted to the matching between the 2HDM and the SMEFT; we discuss how decoupling can be used to perform a consistent expansion, and present details of both the tree-level matching up to dimension-8 operators and the one-loop matching.
We present our numerical results in Section~\ref{sec:res}, discussing fits for electroweak precision observables (EWPOs) first, and to Higgs signal strengths  afterwards. We compare numerical results obtained using various SMEFT and loop expansions.
Section~\ref{sec:conclusions} contains some conclusions along with a discussion of future directions for study.
Some appendices complement the main text: Appendix \ref{app:zs} contains formulae in the 2HDM, Appendix \ref{app:loop matching} contains the one-loop matching results relevant for our analyses, and Appendix \ref{app:Higgs} provides details on the fits. 
Details of our 2HDM implementation of \texttt{Matchmakereft} and \texttt{Matchete} can be found in the auxiliary material \cite{auxiliary:2024bdfh}.

\section{2HDM}
\label{sec:twhdm}

\subsection{The model}
\label{sec:model}

For this review of the 2HDM, we follow Ref.~\cite{Dawson:2022cmu} closely (for more details, cf. Refs.~\cite{Gunion:2002zf, Gunion:1989we, Branco:2011iw}). The model contains a second scalar doublet $\Phi_2$ along with the SM scalar doublet $\Phi_1$, with real vacuum expectation values (vevs) $v_2/\sqrt{2}$ and $v_1/\sqrt{2}$, respectively. A softly broken $Z_2$ symmetry is imposed on the potential, under which the scalar doublets transform as $\Phi_1 \to \Phi_1$ and $\Phi_2 \to -\Phi_2$. It is convenient to rotate the fields to the Higgs basis, with doublets $H_1$ and $H_2$:
\bea
\label{eq:basis-rot}
\left(\begin{array}{c}
H_{1} \\
H_{2}
\end{array}\right)=\left(\begin{array}{cc}
c_{\beta} & s_{\beta} \\
-s_{\beta} & c_{\beta}
\end{array}\right)\left(\begin{array}{c}
\Phi_{1} \\
\Phi_{2}
\end{array}\right),
\eea
where we introduced the short notation $s_x \equiv \sin x, c_x \equiv \cos x$, and  $\tan \beta = v_2/v_1$. This implies that, in the Higgs basis, only $H_1$ has a vev, $v/\sqrt{2}$, with $v = \sqrt{v_1^2 + v_2^2}=246$ GeV.

We focus on two terms of the Lagrangian, $\mathcal{L}_{2 \mathrm{HDM}} \supset \mathcal{L}_Y -V$ (representing the Yukawa terms and the potential, respectively), and we write them both in the Higgs basis. 
The $Z_2$ symmetry is extended to the fermions in order to avoid flavor changing neutral currents at tree-level. There are four possibilities for such an extension, leading to four types of 2HDM: Type-I, Type-II, Type-L and Type-F. We write $\mathcal{L}_Y$ 
as:
\bea
\label{eq:Yuk-Lag}
\mathcal{L}_Y = 
- Y_u {\overline q}_L {\tilde {H}} _1 u_R 
- Y_d{\overline q}_L H_1 d_R 
- Y_e {\overline l}_L H_1 e_R 
 - \frac{\eta_u Y_u}{\tan\beta}\, {\overline q}_L {\tilde {H}} _2 u_R 
- \frac{\eta_d Y_d}{\tan\beta}\, {\overline q}_L H_2 d_R
- \frac{\eta_e Y_e}{\tan\beta}\, {\overline l}_L H_2 e_R
+ \text{h.c.} \, ,
\eea
where $\tilde{H}_i=i \sigma_2 H_i^*$ and where we suppress generation indices on the left-handed SU(2)$_L$ doublets $q_L$ and $l_L$, and on the right-handed SU(2)$_L$ singlets $u_R, d_R$ and $e_R$. 
The Yukawa matrices are related to the fermion mass matrices via
$Y_f=\sqrt{2} \, M_f/v$, where $f$ represents any type of fermion: up-type ($u$) and down-type ($d$) quarks and charged leptons ($e$). Here, $M_f$ represents a $3\times 3$ matrix in flavor space, whose singular values are the masses $m_f$ of the fermions of type $f$. Finally, the parameters $\eta_f$ specify the type of 2HDM and are given in Table \ref{tab:types}.

As for the potential, we write it  as: 
\begin{eqnarray}
\label{eq:potential}
V &=& Y_1 H_{1}^{\dagger} H_{1}
+ Y_2 H_{2}^{\dagger} H_{2}+\left(Y_3 H_{1}^{\dagger} H_{2}+\textrm{h.c.}\right) 
\nonumber \\[0.25em]
&&+ \frac{Z_{1}}{2}\left(H_{1}^{\dagger} H_{1}\right)^{2}+\frac{Z_{2}}{2}\left(H_{2}^{\dagger} H_{2}\right)^{2}+Z_{3}\left(H_{1}^{\dagger} H_{1}\right)\left(H_{2}^{\dagger} H_{2}\right)+Z_{4}\left(H_{1}^{\dagger} H_{2}\right)\left(H_{2}^{\dagger} H_{1}\right) \nonumber \\[0.25em]
&& + \left\{\frac{Z_{5}}{2}\left(H_{1}^{\dagger} H_{2}\right)^{2}+Z_{6}\left(H_{1}^{\dagger} H_{1}\right)\left(H_{1}^{\dagger} H_{2}\right)+Z_{7}\left(H_{2}^{\dagger} H_{2}\right)\left(H_{1}^{\dagger} H_{2}\right)+ \textrm{h.c.}\right\}.
\end{eqnarray}
The parameters $Y_3, Z_5, Z_6, Z_7$ are in general complex, whereas the remaining ones are real. In this paper, we assume the particular case in which $Y_3, Z_5, Z_6, Z_7$ only take real values.%
\fn{This assumption is usually understood as defining a model, known as the real 2HDM. Ref. \cite{Fontes:2021znm} argues that the real 2HDM might not be a fully
consistent theory.}
This implies CP conservation at the tree-level in the scalar sector, in which case the doublets can be written as
\begin{equation}
H_1 = 
\begin{pmatrix}
G^+ \\
\frac{1}{\sqrt{2}}(v + h_1^{\mathrm{H}} + i G_0)
\end{pmatrix}\qquad 
H_2 = 
\begin{pmatrix}
H^+ \\
\frac{1}{\sqrt{2}}(h_2^{\mathrm{H}} + i A)
\end{pmatrix},
\end{equation}
where $h_1^{\mathrm{H}}, h_2^{\mathrm{H}}, G_0$ and $A$  are real fields and $G^+, H^+$ are complex fields. The mass matrix for $h_1^{\mathrm{H}}$ and $h_2^{\mathrm{H}}$ can be diagonalized with a mixing angle $\alpha$,
\begin{equation}
\label{eq:diagonalization}
\left(\begin{array}{c}
h \\
H
\end{array}\right)
=
\left(\begin{array}{cc}
\sba & \cba\\
\cba & - \sba
\end{array}\right)
\left(\begin{array}{c}
h_1^{\mathrm{H}}\\
h_2^{\mathrm{H}}
\end{array}\right),
\end{equation}
where $h$ and $H$ are the neutral scalar mass states, with $h$ being the $125\,\textrm{GeV}$ scalar that is observed at the LHC.
We assume $0 \leq \beta-\alpha \leq \pi$, so that $\sba = \sqrt{1 - \cba^2} > 0$.
Finally, defining the masses of $h, H, A$ and $H^{+}$ to be $m_{h},~ m_{H},~ m_A$ and $m_{H^{+}}$, respectively, we take the following parameters as independent:
\begin{equation}
\label{eq:indep-real}
\cba,
\,
\beta,
\,
v,
\,
m_f,
\,
m_{h},
\,
Y_2,
\,
m_{H},
\,
m_A,
\,
m_{H^{+}}.
\end{equation}
Expressions for $Y_1$ and $Y_3$, along with the $Z_i$ parameters of Eq. (\ref{eq:potential}), are given in Appendix~\ref{app:zs} in terms of the parameters of \eqref{eq:indep-real}.

\begin{table}[t!]
\renewcommand{\arraystretch}{1.2}
\centering
\begin{tabular}
{
@{\hspace{-0.8mm}}
>{\centering}p{1cm}
>{\centering}p{1.8cm}
>{\centering}p{1.8cm}
>{\centering}p{1.8cm}
>{\centering\arraybackslash}p{1.8cm}
@{\hspace{3mm}}
}
\hlinewd{1.1pt}
%\hline
& Type-I &  Type-II & Type-L & Type-F \\
\hline
$\eta_{u} $ & 1 & 1 & 1 & 1 \\
$\eta_{d}$ & 1 & $-\tan ^{2} \beta$ & 1 & $-\tan ^{2} \beta$ \\
$\eta_{e}$ & 1 & $-\tan ^{2} \beta$ & $-\tan ^{2} \beta$ & 1 \\
%\hline
\hlinewd{1.1pt}
\end{tabular}
\caption{Values of the parameter $\eta_f$ for the different types of 2HDM models and for the different types of charged fermions.}
\label{tab:types}
\end{table}
\normalsize

\subsection{Constraints on the parameters}
\label{sec:lims}

The 2HDM is limited by a number of theoretical constraints which all push the allowed parameters towards the alignment limit, $\cos(\beta-\alpha) = 0$.  These constraints do not involve the fermion couplings (at tree-level) and so apply to all the types of 2HDMs studied here. 
The limits from perturbativity  require that the scalar quartic couplings be less than $4\pi$, while perturbative unitarity of the $2\rightarrow 2$ scalar scattering processes in the high energy limit requires that the eigenvectors of the scattering matrix be less than $8\pi$~\cite{Ginzburg:2005dt}. 
Finally, there is the requirement that the potential be bounded from below~\cite{Ivanov:2015nea}, which is essentially the requirement that the quartic couplings be positive. These constraints, taken together, imply that there is very little allowed parameter space away from the $\cos(\beta-\alpha) \rightarrow 0$ limit (as will be seen later explicitly in Fig. ~\ref{fig:type1-extras}).
 
There are significant experimental constraints on the scalar sector of the 2HDM coming from $B$ meson decays.  For Types-I, L and F, the charged Higgs contribution to $b\rightarrow s\gamma$ requires $\tan\beta > 1.2$, while for Type-II it requires $m_{H^+}> 600\,\textrm{GeV}$ for all values of $\tan\beta$\cite{Haller:2018nnx}. The Type-II model has the further restriction from $B\rightarrow \mu^+\mu^-$ that, for $m_{H^+}\sim 1\,\textrm{TeV}$, we must have $\tan\beta\lsim 25$.

%%%%%%%%%%%%%
\section{Matching}
\label{sec:matching}

In this section, we discuss the matching between the SMEFT and the 2HDM.%
\fn{Concerning the matching between an EFT and the 2HDM, cf. Refs.~\cite{Gorbahn:2015gxa,Brehmer:2015rna,Egana-Ugrinovic:2015vgy,Belusca-Maito:2016dqe,Belusca-Maito:2017iob,Belusca-Maito:2016cay,Banta:2023prj,Dawson:2023ebe,Dawson:2023oce,Arco:2023sac,Buchalla:2023hqk}.}
We start by briefly discussing the procedure in Section~\ref{sec:decoupling}, focusing on the notion of decoupling, and we establish our conventions in Section~\ref{sec:conventions}. We then present the matching equations: in Section~\ref{sec:tree-matching} for the tree-level matching, and in Section~\ref{sec:loop-matching} (and Appendix~\ref{app:loop matching}) for the one-loop matching. Finally, we describe in Section~\ref{sec:SMEFT-relations}
relations between quantities in the SMEFT.

\subsection{Matching with decoupling}
\label{sec:decoupling}

To obtain the matching (at both tree-level and one-loop), one starts with the 2HDM before spontaneous symmetry breaking (SSB). The dimensionful parameter $Y_2$, as defined in the Higgs basis, is assumed to be very large. The heavy degrees of freedom are then integrated out, leading to an effective Lagrangian corresponding to an expansion in inverse powers of $Y_2$. This parameter is thus identified with the SMEFT mass scale squared $\Lambda^2$, and the matching equations establish a relation between the SMEFT coefficients and the parameters of Eqs. (\ref{eq:Yuk-Lag}) and (\ref{eq:potential}). Note that this assumes that $Y_2$ is the only large parameter. 

It is then convenient to rewrite the matching equations in terms of the parameters of \eqref{eq:indep-real}. To that end, one needs to specify how each one of those parameters scale. This can be done resorting to the notion of the decoupling limit of the 2HDM \cite{Haber:1989xc,Gunion:2002zf,Haber:2006ue,Asner:2013psa,Dawson:2023ebe,Buchalla:2023hqk,Arco:2023sac}, which takes the mass states $H$, $A$ and $H^{+}$ to be heavy. This is a reasonable scenario, as the EFT will in general not be applicable if at least one of those states is light
\cite{Belusca-Maito:2016dqe}.
Moreover, it is only when these non-SM states decouple that the SM can be approximated by  an EFT. As we mentioned, the decoupling limit establishes a scaling for the parameters of \eqref{eq:indep-real}; we follow Ref.~\cite{Dawson:2023ebe} to characterize that limit as:
\bs
\label{eq:decoupling}
\bea
\label{eq:decoupling-1}
&Y_2 =\Lambda^2, 
\quad
m_H^2=\Lambda^2+\Delta m_H^2, \quad m_A^2 =\Lambda^2+\Delta m_A^2, \quad m_{H^{+}}^2=\Lambda^2+\Delta m_{H^{+}}^2,& \\
&\Lambda^2 \gg v^2, \quad m_h^2 \sim \mathcal{O}\left(v^2\right), \quad \Delta m_H^2, \Delta m_A^2, \Delta m_{H^{+}}^2 \sim \mathcal{O}\left(v^2\right), \quad \cba \sim \mathcal{O}\left(v^2 / \Lambda^2\right), &
\eea
\es
where the $\Delta m^2$ parameters ($\Delta m_H^2$, $\Delta m_A^2$ and $\Delta m_{H^{+}}^2$) are real, and where the symbol $\sim$ denotes scaling. As in Ref.~\cite{Dawson:2023ebe}, we find it convenient to introduce an auxiliary dimensionless parameter $\xi$. This is a small quantity that we use to organize the expansion, such that $v^2/\Lambda^2 \sim \cba \sim {\cal{O}}(\xi)$, which we implement in our codes by the replacements
\be
\label{eq:scaling-with-xi}
\frac{1}{\Lambda^2}\rightarrow \frac{\xi}{\Lambda^2},
\qquad
\cba\rightarrow \xi \, \cba,
\ee
with all other parameters being of ${\cal{O}}(\xi^0)$.
Therefore, when writing the matching relations between the SMEFT coefficients and the parameters of \eqref{eq:indep-real}, instead of taking into account simply the heavy scale $\Lambda$, we take into account both $\Lambda$ and $\cba$ by performing an expansion in powers of $\xi$. Note that the expansion is only formally consistent if $c_{\beta-\alpha}$ is small (i.e. close to the alignmenent limit, $c_{\beta-\alpha}=0$).

\subsection{Conventions}
\label{sec:conventions}

We write the SMEFT Lagrangian matched to the 2HDM as:
\be
\label{eq:Lag-matched}
\mathcal{L}_{\mathrm{SMEFT}}^{\rm matched} = \mathcal{L}_{\mathrm{SM}} + \mathcal{L}_6^{[t]} + \mathcal{L}_6^{[l]} + \mathcal{L}_8 + \mathcal{O}(\Lambda^{-6}),
\ee
where $\mathcal{L}_6^{[t]}$, $\mathcal{L}_6^{[l]}$ and $\mathcal{L}_8$ 
represent
the set of terms containing dimension-6 operators generated via tree-level matching,
the set containing dimension-6 operators generated via one-loop matching
and the set containing dimension-8 operators generated via tree-level matching, respectively.%
\fn{We do not consider dimension-8 operators generated via one-loop matching\cite{Banerjee:2023iiv,Banerjee:2023xak}.} 
We follow the Warsaw basis conventions~\cite{Grzadkowski:2010es} for dimension-6 operators and those of Murphy~\cite{Murphy:2020rsh} for dimension-8 operators.
We also use the superscripts $[t]$ and $[l]$ for the SMEFT coefficients of dimension-6 operators of the Warsaw basis which are generated in both the tree-level and the one-loop matchings;\fn{We use superscripts $[t]$ and $[l]$ instead of the conventional (0) and (1) to prevent any potential confusion with the names of the SMEFT WCs.} if $O_x$ is one such operator, we have:
\be
\label{eq_t-vs-l}
\mathcal{L}_{\mathrm{SMEFT}}^{\rm matched} \supset
\dfrac{C_x O_x}{\Lambda^2} = \dfrac{C_x^{[t]} O_x}{\Lambda^2} + \dfrac{C_x^{[l]} O_x}{\Lambda^2}.
\ee
In this way, $C_x^{[t]}$ ($C_x^{[l]}$) represents the component of the dimension-6 SMEFT coefficient generated via tree-level (one-loop) matching and is included in $\mathcal{L}_6^{[t]}$ ($\mathcal{L}_6^{[l]}$).
We present $\mathcal{L}_6^{[t]}$ and $\mathcal{L}_8$ in Section~\ref{sec:tree-matching}, leaving $\mathcal{L}_6^{[l]}$ to Section~\ref{sec:loop-matching}.

Besides the parameters of \eqref{eq:indep-real} (subject to \eqref{eq:decoupling-1}), we take as our input parameters $m_W$, $m_Z$, and $G_F$ (representing the $W$-boson mass, the $Z$-boson mass and the Fermi constant, respectively), and give all results in terms of these parameters. 
We define $\phi$ as the SMEFT Higgs doublet.
In our results, we consider terms up to $\mathcal{O}(\xi^2)$. We assume loop factors to be of the same order as $\mathcal{O}(\xi)$, which means we consistently neglect loop  generated terms which are $ \mathcal{O}(\xi^2)$.

By default, we write the WCs and the fermion operators in a generation-independent way. Whenever it is relevant to specify generations, we do it by writing them in a subscript, separated from any previous subscript by a comma. In the tree-level matching, we do not write the contributions from operators with leptons (which are trivially obtained from those with down-type quarks), and also omit the contributions from 4-fermion operators (which are not relevant for our analyses). As for the loop matching, the operators 
$O_{\phi l}^{(3)} \equiv \Big(\phi^{\dagger} i \lrDmuI \phi \Big) \left(\bar{l}_L \tau^I \gamma^\mu l_L \right)$
and
$O_{l l} = \left(\bar{l}_L \gamma_\nu l_L \right) \left(\bar{l}_L \gamma^\nu l_L \right)$
are generated, and contribute to the relation between $G_F$ and the SMEFT vev.%
\fn{$O_{l l}$ is also generated via tree-level matching, but it is neglected there as it multiplies masses of light leptons. This is not the case in the loop matching, so that we keep the operator in that case. The $\tau$'s represent Pauli matrices.}
We assume flavor universality in the generations involved; accordingly, we define operators with bold subscripts such that:
\begin{equation}
\begin{aligned}
& C_{\phi l,11}^{(3)} \Big(\phi^{\dagger} i \lrDmuI \phi \Big)\left(\bar{l}_{L,1} \tau^I \gamma^\mu l_{L,1}\right)
+
C_{\phi l,22}^{(3)} \Big(\phi^{\dagger} i \lrDmuI \phi \Big)\left(\bar{l}_{L,2} \tau^I \gamma^\mu l_{L,2}\right) \\
&
% \qquad \,\, = \,\,
% 2 \, C_{\phi l,11}^{(3)} \Big(\phi^{\dagger} i \lrDmuI \phi\Big)\left(\bar{l}_{L,1} \tau^I \gamma^\mu l_{L,1}\right) 
% \\
% %
% &\qquad 
\,\, \equiv \,\, 2 \, C_{\bm{\phi l}}^{(3)} \Big(\phi^{\dagger} i \lrDmuI \phi\Big)\left(\bar{l}_L \tau^I \gamma^\mu l_L\right)\,,
\end{aligned}
\end{equation}
and
\begin{equation}
\begin{aligned}
\label{eq:Oll}
& C_{ll,1221} \left(\bar{l}_{L,1} \gamma_\nu l_{L,2} \right) \left(\bar{l}_{L,2} \gamma^\nu l_{L,1} \right) + \mathrm{h.c.} \\
&\qquad \, \, = \, \, C_{ll,1221} \left(\bar{l}_{L,1} \gamma_\nu l_{L,2} \right) \left(\bar{l}_{L,2} \gamma^\nu l_{L,1} \right) 
+ C_{ll,2112} \left(\bar{l}_{L,2} \gamma_\nu l_{L,1} \right) \left(\bar{l}_{L,1} \gamma^\nu l_{L,2} \right) \\ 
&\qquad
% \,\, =\,\, 2 \, C_{ll,1221} \left(\bar{l}_L \gamma_\nu l_L \right) \left(\bar{l}_L \gamma^\nu l_L \right)
\,\, \equiv \,\,
\, 2 \, C_{\bm{ll}} \left(\bar{l}_L \gamma_\nu l_L \right) \left(\bar{l}_L \gamma^\nu l_L \right)\, ,
\end{aligned}
\end{equation}
where in the last equalities we take only a single choice of generation indices and do not sum over them.

\subsection{Tree-level matching}
\label{sec:tree-matching}

The tree-level matching relations up to dimension-8 SMEFT operators were obtained in Ref.~\cite{Dawson:2022cmu}. However, that reference considered the case $\Delta m_H^2 = \Delta m_A^2 = \Delta m_{H^{+}}^2 = 0$. Here we consider the general case where these parameters can be non-zero.
The dimension-6 Lagrangian, $\mathcal{L}_6^{[t]}$, of \eqref{eq:Lag-matched} is \cite{Dawson:2022cmu}:
\be
\label{eq:L6-tree}
\mathcal{L}_6^{[t]} = 
\frac{C_{\phi}^{[t]}}{\Lambda^2} \, (\phi^{\dagger} \phi)^3
+ 
\left\{
\frac{C_{u\phi}^{[t]}}{\Lambda^2} \, (\phi^{\dagger} \phi) \, \bar{q}_L u_R \tilde{\phi} +
\frac{C_{d\phi}^{[t]}}{\Lambda^2} \, (\phi^{\dagger} \phi) \, \bar{q}_L d_R \phi +
\mathrm{h.c.} \right\} + \mathrm{4F},
\ee
where 4F represents 4-fermion operators.%
\fn{As discussed above, we omit the contributions from operators with leptons in the tree-level matching; they are trivially obtained from those with down-type quarks.}
The matching equations are:
\bs
\label{eq:WC-dim6-tree}
\bea
\dfrac{C_{\phi}^{[t]}}{\Lambda^2} \mym &=& \mym 2 G_F^2 c^2_{\beta-\alpha} \Lambda^2 + 8 G_F^2 c^2_{\beta-\alpha} (\Delta m_H^2 - m_h^2), \\
\label{eq:c-u-phi-tree}
\dfrac{C_{u\phi}^{[t]}}{\Lambda^2} \mym &=& \mym - \dfrac{2}{\sqrt{2}} 
(\sqrt{2} G_F)^{3/2} \dfrac{\cba}{\tan\beta} \eta_u {m_u} 
- \dfrac{\cba {m_u} (\sqrt{2} G_F)^{3/2}}{\sqrt{2} {\Lambda}^2}
\bigg[ \cba \, {\Lambda}^2 + \dfrac{2 \eta_u}{\tan\beta} (2 \Delta m_H^2 - 3 m_h^2)\bigg], \\
\label{eq:c-d-phi-tree}
\dfrac{C_{d\phi}^{[t]}}{\Lambda^2} \mym &=& \mym - \dfrac{2}{\sqrt{2}} 
(\sqrt{2} G_F)^{3/2} \dfrac{\cba}{\tan\beta} \eta_d {m_d} 
- \dfrac{\cba {m_d} (\sqrt{2} G_F)^{3/2}}{\sqrt{2} {\Lambda}^2} \bigg[ \cba \, {\Lambda}^2 + \dfrac{2 \eta_d}{\tan\beta} (2 \Delta m_H^2 - 3 m_h^2) \bigg].
\eea
\es
where $m_f$ represents the mass of fermion of type $f$ (recall section \ref{sec:model}).
We see that, besides the 4-fermion operators, only 2 kinds of dimension-6 operators are generated, $O_{\phi}$ and $O_{f\phi}$. In particular, the dimension-6  SMEFT matched  at tree-level with the 2HDM has no information about the 2HDM interaction between the Higgs and gauge bosons. As for the Yukawa interactions, the coefficients of $O_{f\phi}$ are proportional to $m_f$ and depend on the type of 2HDM via the parameters $\eta_f$. We also note that, even though the WCs of \eqref{eq:WC-dim6-tree} are generated immediately at $\mathcal{O}(\xi^1)$, they have $\mathcal{O}(\xi^2)$ corrections (which can be seen by comparing the expressions of that equation with Eqs.~(\ref{eq:decoupling}) and (\ref{eq:scaling-with-xi})); this happens in such a way that it is only at $\mathcal{O}(\xi^2)$ that the $\Delta m^2$ corrections show up. Finally, the WCs of \eqref{eq:WC-dim6-tree} are computed at the scale $\Lambda$. Renormalization group evolution (RGE) can be used to evolve the coefficients to the weak scale \cite{Jenkins:2013zja,Jenkins:2013wua,Alonso:2013hga}. As mentioned above, we consistently work to $\mathcal{O}(\xi^2)$, and assume that loop factors are equivalent to an additional factor of $\mathcal{O}(\xi)$. This means that only the RGE of the  $\mathcal{O}(\xi^1)$ terms of Eq. (\ref{eq:WC-dim6-tree}) are included (in Fig.~\ref{fig:type1-extras}, we will demonstrate that this effect is numerically small).

The dimension-8 Lagrangian, $\mathcal{L}_8$, of \eqref{eq:Lag-matched} is \cite{Dawson:2022cmu}:
\bea
\label{eq:L8}
\mathcal{L}_8 \mym &=& \mym \frac{C_{\phi^8}}{\Lambda^4} (\phi^{\dagger} \phi)^4
+
\frac{C_{\phi^6}^{(1)}}{\Lambda^4} (\phi^{\dagger} \phi)^2 \left(D_{\mu} \phi\right)^{\dagger} \left(D^{\mu} \phi\right)
+
\bigg\{
\frac{C_{qu\phi^5}}{\Lambda^4} (\phi^{\dagger} \phi)^2 \bar{q}_L u_R \tilde{\phi} +
\frac{C_{qu\phi^3D^2}^{(1)}}{\Lambda^4} (D_{\mu} \phi)^{\dagger} (D^{\mu} \phi) \bar{q}_L u_R \tilde{\phi} \no
&&
+ \frac{C_{qu\phi^3D^2}^{(2)}}{\Lambda^4} \Big[(D_{\mu} \phi)^{\dagger} \tau^I (D^{\mu} \phi) \Big] \Big[\bar{q}_L u_R \tau^I \tilde{\phi}\Big] +
\frac{C_{qu\phi^3D^2}^{(5)}}{\Lambda^4} \Big[(D_{\mu}\phi)^{\dagger} \phi \Big] \Big[ \bar{q}_L u_R \widetilde{D^{\mu} \phi} \Big] + C_{qd\phi^5} (\phi^{\dagger} \phi)^2 \bar{q}_L d_R \phi  \no
&&
+ \frac{C_{qd\phi^3D^2}^{(1)}}{\Lambda^4} (D_{\mu} \phi)^{\dagger} (D^{\mu} \phi) \bar{q}_L d_R \phi + \frac{C_{qd\phi^3D^2}^{(2)}}{\Lambda^4} \Big[(D_{\mu} \phi)^{\dagger} \tau^I (D^{\mu} \phi) \Big] \Big[\bar{q}_L d_R \tau^I \phi\Big]  \no
&&
+ \frac{C_{qd\phi^3D^2}^{(5)}}{\Lambda^4} (\phi^{\dagger} D_{\mu} \phi) (\bar{q}_L d_R D^{\mu} \phi) + \mathrm{h.c.} \bigg\} + \mathrm{4F}, 
\eea
with:
\bs
\label{eq:more-WCs-after-decoupling}
\bea
\frac{C_{\phi^8}}{\Lambda^4} &=& 2 \cba^2 (\sqrt{2} G_F)^3 (m_h^2 - \Delta m_H^2)
, \\
\label{eq:46b}
\frac{C_{\phi^6}^{(1)}}{\Lambda^4} &=& - \cba^2 \, (\sqrt{2} G_F)^2 
, \\
\frac{C_{qu\phi^5}}{\Lambda^4} &=& \dfrac{\sqrt{2} \, \cba \, {m_u} \, (\sqrt{2} G_F)^{5/2}}{{\Lambda}^2 } \bigg[\cba \, {\Lambda}^2 + \dfrac{\eta_u}{\tan\beta} (2 \Delta m_H^2 - 3 m_h^2) \bigg]
, \\
\frac{C_{qu\phi^3D^2}^{(1)}}{\Lambda^4} &=& \dfrac{3 \, \sqrt{2} \, \cba \, {m_u} \, (\sqrt{2} G_F)^{3/2}  \, \eta_u}{{\tan \beta \, \Lambda}^2}
, \\
\frac{C_{qu\phi^3D^2}^{(2)}}{\Lambda^4} &=& - \dfrac{\sqrt{2} \, \cba \, {m_u} \, (\sqrt{2} G_F)^{3/2}  \, \eta_u}{{\tan \beta \, \Lambda}^2 }
, \\
\frac{C_{qu\phi^3D^2}^{(5)}}{\Lambda^4} &=& \dfrac{2 \, \sqrt{2} \, \cba \, {m_u} \, (\sqrt{2} G_F)^{3/2}  \, \eta_u}{{\tan \beta \, \Lambda}^2 }
, \\
\frac{C_{qd\phi^5}}{\Lambda^4} &=& \dfrac{\sqrt{2} \cba {m_d} (\sqrt{2} G_F)^{5/2}}{\Lambda^2 } \bigg[ \cba {\Lambda}^2 + \dfrac{\eta_d}{\tan\beta} (2 \Delta m_H^2 - 3 m_h^2) \bigg]
, \\
\frac{C_{qd\phi^3D^2}^{(1)}}{\Lambda^4} &=& \dfrac{3 \, \sqrt{2} \, \cba \, {m_d} \, (\sqrt{2} G_F)^{3/2}  \, \eta_d}{{\tan \beta \, \Lambda}^2 }
, \\
\frac{C_{qd\phi^3D^2}^{(2)}}{\Lambda^4} &=& \dfrac{\sqrt{2} \, \cba \, {m_d} \, (\sqrt{2} G_F)^{3/2}  \, \eta_d}{{\tan \beta \, \Lambda}^2 }
, \\
\frac{C_{qd\phi^3D^2}^{(5)}}{\Lambda^4} &=& \dfrac{2 \, \sqrt{2} \, \cba \, {m_d} \, (\sqrt{2} G_F)^{3/2}  \, \eta_d}{{\tan \beta \, \Lambda}^2 }\, .
\eea
\es
Note that the number of dimension-8 operators which are generated at tree-level and which are relevant for our analyses is not very large. Information about the 2HDM interactions between the Higgs and gauge bosons appears via the operator $O_{{\phi}^6}^{(1)}$. As in the case of dimension-6 operators, the only $\Delta m^2$ corrections that contribute are those of $\Delta m_H^2$.

\subsection{One-loop matching}
\label{sec:loop-matching}

The complete one-loop matching between the 2HDM and the SMEFT with dimension-6 operators is presented here for the first time. Partial results were already derived in Ref.~\cite{Anisha:2021hgc}; we checked that our results are consistent with those of that reference. 
We use the codes \texttt{Matchmakereft} and \texttt{Matchete}. Both software packages yield their results in the Green's basis of Ref.~\cite{Carmona:2021xtq} (up to integration by parts and Fierz relations), such that the comparison can be easily done in this basis. We confirm that the results agree. \texttt{Matchmakereft} also provides the results in the Warsaw basis \cite{Grzadkowski:2010es} (which is the basis we follow in this paper to write the operators of dimension-6). The full results (in both bases) are contained in the auxiliary files \cite{auxiliary:2024bdfh}. Here, we show only the operators in the Warsaw basis that contribute to electroweak precision observables at LO in the SMEFT \cite{Dawson:2019clf}, as well as to the leading contributions to  Higgs production and decay:%
\fn{As usual, we omit generation indices, and we follow \eqref{eq_t-vs-l} by including the superscript $[l]$ in those dimension-6 WCs that are also generated via tree-level matching.}
\bea
\label{eq:L6-loop}
\mathcal{L}_6^{[l]} \mym &\supset& \mym
\frac{C_{\phi W}}{\Lambda^2}\, (\phi^{\dagger} \phi) W_{\mu \nu}^I W^{I \mu \nu}
+
\frac{C_{\phi B}}{\Lambda^2} \, (\phi^{\dagger} \phi) B_{\mu \nu} B^{\mu \nu}
+
\frac{C_{\phi WB}}{\Lambda^2} \, (\phi^{\dagger} \tau^I \phi) W_{\mu \nu}^I B^{\mu \nu} \no
&&
+
\frac{C_{\phi}^{[l]}}{\Lambda^2} \, (\phi^{\dagger} \phi)^3
+
\frac{C_{\phi \Box}}{\Lambda^2} \, (\phi^{\dagger} \phi) \square(\phi^{\dagger} \phi)
+
\frac{C_{\phi D}}{\Lambda^2} \left(\phi^{\dagger} D^\mu \phi\right)^{\star}\left(\phi^{\dagger} D_\mu \phi\right) \no
&&
%\hspace{-3mm}
+
 \frac{C_{\phi u}}{\Lambda^2} \left(\phi^{\dagger} i \lrDmu \phi\right)\left(\bar{u}_R \gamma^\mu u_R \right)
+
\frac{C_{\phi d}}{\Lambda^2} \left(\phi^{\dagger} i \lrDmu \phi\right)\left(\bar{d}_R \gamma^\mu d_R \right)
+
\frac{C_{\phi e}}{\Lambda^2} \left(\phi^{\dagger} i \lrDmu \phi\right)\left(\bar{e}_R \gamma^\mu e_R \right) \no
&& 
+
\frac{C_{\phi q}^{(1)}}{\Lambda^2} \left(\phi^{\dagger} i \lrDmu \phi\right)\left(\bar{q}_L \gamma^\mu q_L \right)
+
\frac{C_{\phi q}^{(3)}}{\Lambda^2} \left(\phi^{\dagger} i \lrDmuI \phi\right)\left(\bar{q}_L \tau^I \gamma^\mu q_L \right) \no
&& 
+
\frac{C_{\phi l}^{(1)}}{\Lambda^2} \left(\phi^{\dagger} i \lrDmu \phi\right)\left(\bar{l}_L \gamma^\mu l_L \right) +
\frac{C_{\phi l}^{(3)}}{\Lambda^2} \left(\phi^{\dagger} i \lrDmuI \phi\right)\left(\bar{l}_L \tau^I \gamma^\mu l_L \right)
+
\frac{C_{\bm{ll}}}{\Lambda^2} \left(\bar{l}_L \gamma_\mu l_L \right) \left(\bar{l}_L \gamma^\mu l_L \right) \no
&&
+ 
\bigg\{
\frac{C_{u\phi}^{[l]}}{\Lambda^2} \, (\phi^{\dagger} \phi) \, \bar{q}_L u_R \tilde{\phi} 
+
\frac{C_{d\phi}^{[l]}}{\Lambda^2} \, (\phi^{\dagger} \phi) \, \bar{q}_L d_R \phi
+
\frac{C_{e \phi}^{[l]}}{\Lambda^2} \, (\phi^{\dagger} \phi) \, \bar{l}_L e_R \phi 
+ \mathrm{h.c.} \bigg\}. \hspace{5mm}
\eea
The one-loop matching contributions to these coefficients are presented in Appendix \ref{app:loop matching}.

\subsection{SMEFT Relations}
\label{sec:SMEFT-relations}

Now that all the terms in \eqref{eq:Lag-matched} have been specified, there are only two tasks required to perform calculations in the SMEFT (more specifically, in the SMEFT matched to the 2HDM): to write the relevant dependent parameters in terms of the independent ones, and to ensure that the fields are canonically normalized. These two tasks were already described in detail in Ref.~\cite{Dawson:2022cmu}, but considering only tree-level matching. Here we extend that discussion to include loop-generated SMEFT dimension-6 operators. Note that we only include the SMEFT operators which are relevant to our analyses, and which were discussed in Sections \ref{sec:tree-matching} and \ref{sec:loop-matching}.

The relevant dependent parameters are those occuring in our calculations, namely ${\overline {g}}_2, v_T$ and $Y_f$. Here, ${\overline {g}}_2$ is the SU(2) gauge coupling occuring in $\mathcal{L}_{\mathrm{SMEFT}}^{\rm matched}$, while $v_T$ is the vev contained in the SMEFT Higgs doublet,
\be
\label{eq:SM-doublet}
\phi
= 
\begin{pmatrix}
G^+_{\text{\tiny S}} \\
\frac{1}{\sqrt{2}}(v_T + h_{\text{\tiny S}} + i G_{0,\text{\tiny S}})
\end{pmatrix},
\ee
with $h_{\text{\tiny S}}$, $G_{0,\text{\tiny S}}$ and $G^+_{\text{\tiny S}}$ being the SMEFT Higgs field, and neutral and charged would-be Goldstone bosons, respectively.
We determine ${\overline {g}}_2$ and $v_T$ from  muon decay and from the expression for the mass of the $W$-boson in $\mathcal{L}_{\mathrm{SMEFT}}^{\rm matched}$.
We find \cite{Dawson:2022cmu,Asteriadis:2022ras}:
\bea
\label{eq:vev}
v_T = 
\dfrac{1}{(\sqrt{2} G_F)^{1/2}}
+
\dfrac{2 \, {C_{\bm{\phi l}}^{(3)}} - {C_{\bm{ll}}}}{2 \, {\Lambda}^2 \, (\sqrt{2} G_F)^{3/2}}
+
\dfrac{1}{8 \, (\sqrt{2} G_F)^{5/2} \Lambda^4}
\left[ 16 \Big(C_{\bm{\phi l}}^{(3)}\Big)^2 - 12 \, {C_{\bm{\phi l}}^{(3)}} \, {C_{\bm{ll}}} + 3 \, (C_{\bm{ll}})^2 - C_{\phi^6}^{(1)} \right]
\eea
and
\be
\label{eq:g2}
{\overline {g}}_2 = 2 \dfrac{m_W}{v_T} = 2 \, m_W \, (\sqrt{2} G_F)^{1/2}
+
\dfrac{m_W \left( {C_{\bm{ll}}} -2 \, {C_{\bm{\phi l}}^{(3)}} \right)}{{\Lambda}^2 (\sqrt{2} G_F)^{1/2}}
+
\dfrac{m_W \left( {C_{\phi ^6}^{(1)}} - 8 \, ({C_{\bm{\phi l}}^{(3)}})^2 + 4 {C_{\bm{\phi l}}^{(3)}} C_{\bm{ll}} - (C_{\bm{ll}})^2\right)}{4 {\Lambda}^4 \, (\sqrt{2} G_F)^{3/2}}.
\ee
Finally, $Y_f$ is determined from the fermion mass in $\mathcal{L}_{\mathrm{SMEFT}}^{\rm matched}$. For the up-type quarks, we find \cite{Dawson:2022cmu,Asteriadis:2022ras}:%
\fn{The expressions for $Y_d$ and $Y_e$ are found trivially from \eqref{eq:yup}.}
\bea
\label{eq:yup}
Y_u \mym &=& \mym
\sqrt{2} \, {M_u} \, (\sqrt{2} G_F)^{1/2} 
+
\dfrac{1}{2 \, {\Lambda}^2 \, (\sqrt{2} G_F)} \bigg[C_{u\phi} + \sqrt{2} \, {M_u} \, (\sqrt{2} G_F)^{1/2} \left(C_{\bm{ll}} - 2 \, {C_{\bm{\phi l}}^{(3)}} \right)   \bigg] \no
&& +
\dfrac{1}{8 \, {\Lambda}^4 \, (\sqrt{2} G_F)^2} \Bigg\{2 \, C_{qu\phi^5} + 8 \, {C_{\bm{\phi l}}^{(3)}} \, C_{u\phi} - 4 \, {C_{\bm{ll}}} \, C_{u\phi}  \no
&& \hspace{35mm} + \sqrt{2} \, {M_u} \, (\sqrt{2} G_F)^{1/2} \left[ {C_{\phi^6}^{(1)}} - 8 \, \Big(C_{\bm{\phi l}}^{(3)}\Big)^2 + 4 \, {C_{\bm{\phi l}}^{(3)}} \, {C_{\bm{ll}}} - (C_{\bm{ll}})^2 \right]   \Bigg\}.
\eea

Concerning the second task referred to above, $h_{\text{\tiny S}}$ does not have a canonically normalized kinetic term. To fix this, we define the Higgs field $h$ with canonically normalized kinetic term, such that:
\be
h_{\text{\tiny S}} = h\, \bigg\{1 + \dfrac{v_T^2}{4 \, {\Lambda}^2} \left( 4 \, {C_{\phi \Box}} - {C_{\phi D}} \right) + \dfrac{v_T^4}{32 \, {\Lambda}^4} \left[ 3 \,  \left( {C_{\phi D}} -4 \, {C_{\phi \Box}} \right) ^2 - 4 \, {C_{\phi ^6}^{(1)}} \right] \bigg\},
\ee
which can be rewritten using \eqref{eq:vev} as:
\bea
h_{\text{\tiny S}} \mym &=& \mym h\, 
\bigg\{
1
+
\dfrac{1}{4 \, \sqrt{2} \, G_F \, {\Lambda}^2} \left(4 \, {C_{\phi \Box}} - {C_{\phi D}} \right) 
+
\dfrac{1}{64 \, G_F^2 \, {\Lambda}^4} \bigg[-4 \, {C_{\phi ^6}^{(1)}} \no
&& \hspace{30mm} + \left( 4 \, {C_{\phi \Box}} - {C_{\phi D}} \right)  \,  \left( 12 \, {C_{\phi \Box}} - 3 \, {C_{\phi D}} + 16 \, {C_{\bm{\phi l}}^{(3)}} - 8 \, {C_{\bm{ll}}} \right) \bigg] 
\bigg\}.
\eea

\section{Numerical results}
\label{sec:res}

In this section, we present our numerical results. We start by discussing fits to EWPOs in Section~\ref{sec:EWPOs}, after which we present fits to Higgs observables in Section~\ref{sec:Higgs}. Our fits are performed not only in the context of the full 2HDM, but also in that of the SMEFT matched to the 2HDM. This allows us to compare both these approaches with the experimental results. More relevant for our purposes, it also allows us to compare the two approaches with one another, and thus investigate the quality with which the different SMEFT truncations replicate the full 2HDM results.

\subsection{Electroweak precision observables}
\label{sec:EWPOs}

The observables related to precision electroweak data that  we consider (defined e.g. in Ref.~\cite{ParticleDataGroup:2022pth}) are
\begin{equation}
m_W,~~
\Gamma_W,~~
\Gamma_Z,~~
\sigma_h,~~
R_e,~~
A_{\textrm{FB},l},~~
R_b,~~
R_c,~~
A_{\textrm{FB},b},~~
A_{\textrm{FB},c},~~
A_b,~~
A_c,~~
A_e\, .
\label{eq:quan}
\end{equation} 
To study the numerical effects for the SMEFT matched to the  full 2HDM,  we start by using the experimental results and the most accurately available SM predictions for the observables of \eqref{eq:quan} that are given in Table III of Ref. \cite{Bellafronte:2023amz}. We then determine the allowed deviations from the SM predictions in the SMEFT, using the 2HDM matched results of Section~\ref{sec:matching} and calculating observables to leading order (LO) in the SMEFT.

There are ten dimension-6 operators (defined in Ref.~\cite{Grzadkowski:2010es}) contributing to the observables of \eqref{eq:quan} at leading order~\cite{Dawson:2019clf}:
\begin{equation}
C_{ll},~~
C_{\phi W B},~~
C_{\phi D},~~
C_{\phi e},~~
C_{\phi u},~~
C_{\phi d},~~
C_{\phi q}^{(1)},~~
C_{\phi q}^{(3)},~~
C_{\phi l}^{(1)},~~
C_{\phi l}^{(3)} \, .
\label{eq:ewpoops}
\end{equation}
When we consider the particular case of the SMEFT matched to the 2HDM, we see from Section~\ref{sec:matching} that none of the dimension-6 operators generated via tree-level matching (\eqref{eq:L6-tree}) coincide with those of \eqref{eq:ewpoops}.%
\fn{Recall that we neglect $C_{ll}$ generated via tree-level matching, since its contribution to EWPOs is suppressed by a small lepton mass.}
Moreover, one can also show that none of the dimension-8 operators generated via tree-level matching (\eqref{eq:L8}) contribute to EWPOs at the leading order.
We conclude that there is no contribution to the EWPOs at leading order if we restrict ourselves to the operators of the SMEFT generated via tree-level matching from the 2HDM.

The situation changes when loop matching is considered. In this case, as can be seen in \eqref{eq:L6-loop}, \textit{all} of the dimension-6 operators in \eqref{eq:ewpoops} are generated, so we expect electroweak precision data to constrain the parameters of the underlying model. 
By considering the expressions for the corresponding WCs (Appendix \ref{app:loop matching}), however, an immediate observation is that none of WCs in \eqref{eq:ewpoops} depend on $\cos(\beta-\alpha)$. 
While they do depend on $\tan\beta$, we will see that the primary dependence is on the $\Delta m^2$.

\begin{figure}[htb!]
\centering
\includegraphics[height=7.8cm]{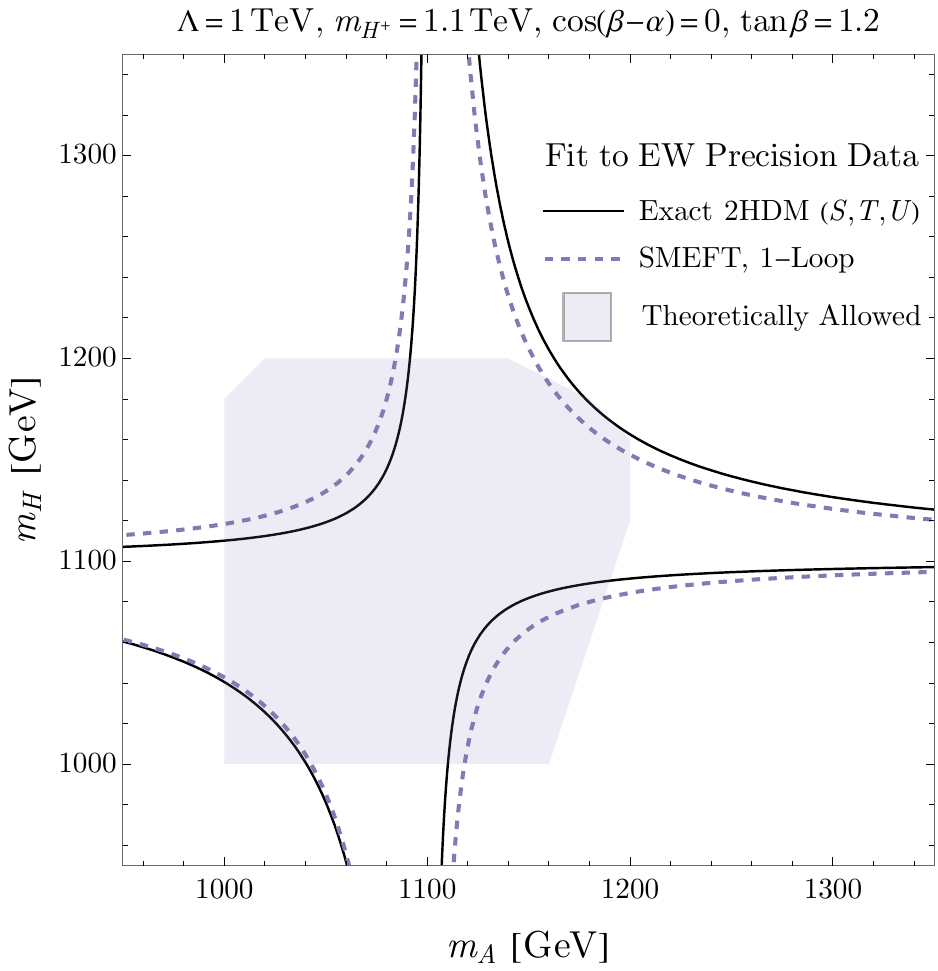}
~
\includegraphics[height=7.8cm]{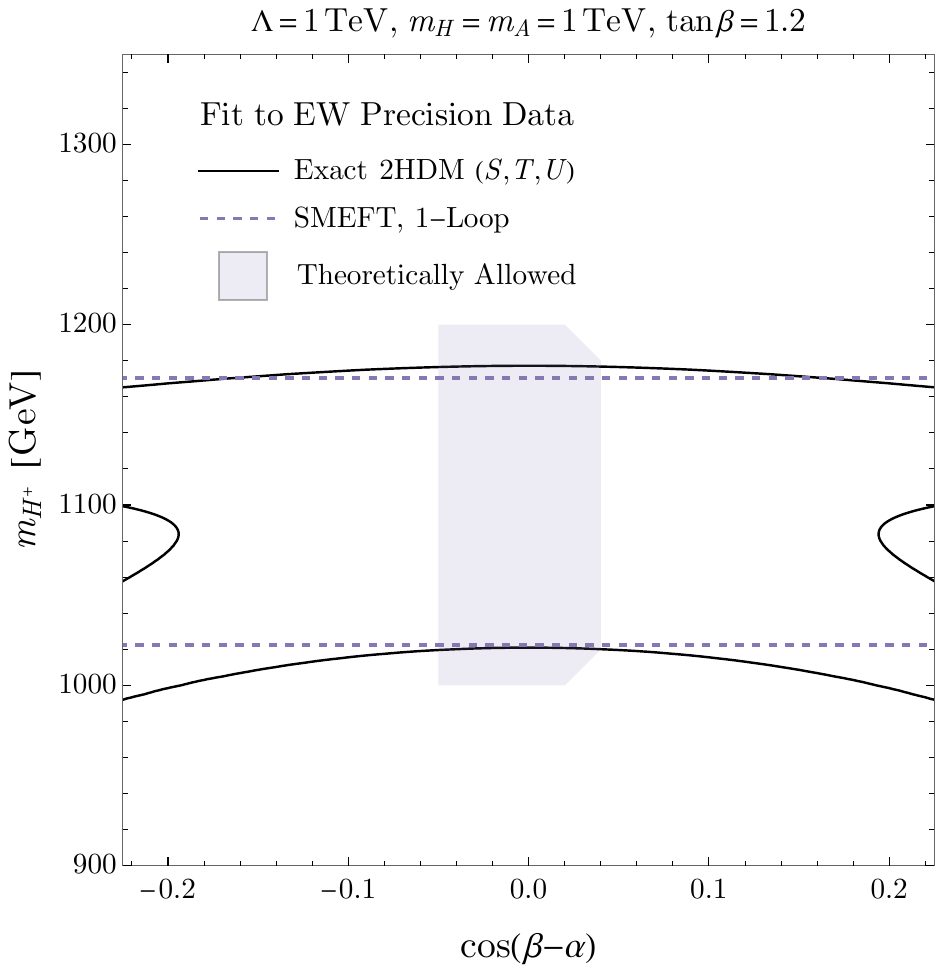}
\hfill
\caption{
Comparison of constraints on the 2HDM from electroweak precision observables in the full 2HDM (via a fit to the oblique parameters $S$, $T$ and $U$) and in the SMEFT matched to the 2HDM (with loop matching). The region satisfying the theoretical constraints outlined in Section~\ref{sec:lims} is shaded gray.}\label{fig:ewpo}
\end{figure}

To compare the SMEFT matched to the 2HDM fits with the limits from EWPO obtained in the full 2HDM, it is sufficient to compute the limits in the full model using the oblique parameters $S$, $T$ and $U$~\cite{Peskin:1991sw}.
Fig.~\ref{fig:ewpo} shows fits to the EWPOs of \eqref{eq:quan} in the full 2HDM and in the SMEFT matched to the 2HDM (with loop-matching).
The left panel considers the alignment limit, $\cos(\beta-\alpha)=0$, and shows an excellent agreement between the full 2HDM results and the results for the SMEFT matched to the 2HDM. The right panel shows $\cos(\beta-\alpha)$ on the horizontal axis. Here, it is again clear that the SMEFT matched to the 2HDM accurately replicates the full 2HDM around the alignment limit. Away from this limit, however, the full 2HDM result changes considerably, whereas the SMEFT matched to the 2HDM does not (as it does not depend on $\cos(\beta-\alpha)$). Finally, both panels show that the SMEFT with one-loop matching provides an excellent description of the full 2HDM in the regions allowed by the theoretical constraints, and for $\cos(\beta-\alpha) \lesssim 0.2$.

In sum, the EWPOs in the 2HDM constitute an example where the SMEFT matched with one-loop dimension-6 operators is clearly more able to accurately replicate the full model than the SMEFT matched with tree-level dimension-8 operators. In fact, whereas the latter do not give a contribution to EWPOs, the former do. On the other hand, this non-zero contribution has no dependence on $\cos(\beta-\alpha)$. This implies that it is adequate for small values of that parameter only, which turn out to be the values preferred by the theoretical constraints.

\subsection{Higgs observables}
\label{sec:Higgs}

We now turn to Higgs observables. We focus on the Higgs signal strengths measured by ATLAS and CMS.
This includes the combined measurements at $\sqrt{s} = 7$ and $8\,\textrm{TeV}$ \cite{ATLAS:2016neq}, as well as the recent ATLAS and CMS combinations at $\sqrt{s} = 13\,\textrm{TeV}$ \cite{ATLAS:2022vkf, CMS:2022dwd}. Details on the fits can be found in Appendix \ref{app:Higgs}.

We start by recapping the results of Ref.~\cite{Dawson:2022cmu}. Ignoring for now the indirect effects of the Higgs self-coupling, the only dimension-6 WCs generated via tree-level matching that are relevant for Higgs observables are $C_{f \phi}$ (recall Eq.~(\ref{eq:L6-tree}) above). In particular, there is no WC contributing to the Higgs couplings to gauge bosons. Now, considering the first term of the r.h.s of Eqs.~(\ref{eq:c-u-phi-tree}) and (\ref{eq:c-d-phi-tree}), as well as Table \ref{tab:types}, it is clear that in the Type-I 2HDM, all contributions to $C_{f \phi}$ at $\mathcal{O}(\xi^1)$ scale as $1/\tan \beta$. 
The dimension-8 operators, however, include information about Higgs-gauge couplings and $\mathcal{O}(\xi^2)$ corrections to the $C_{f\phi}$, both of which are independent of $\tan\beta$. 
Matching up to dimension-8 is thus necessary in the Type-I 2HDM to accurately approximate the full 2HDM for large values of $\tan\beta$.
In contrast, in the other types of 2HDM, there is always at least one type of fermion $f$ for which $C_{f \phi}$ at $\mathcal{O}(\xi^1)$ is not suppressed with $\tan \beta$, so that matching with dimension-8 operators can be neglected in those cases. On the other hand, Ref.~\cite{Dawson:2022cmu} showed that Type-F could accommodate, besides a central region centered around $\cos(\beta-\alpha) =0$, also a disjoint region centered around $\cos(\beta-\alpha) \simeq 0.2$ and large values of $\tan \beta$, usually known as the wrong-sign region.

In what follows, we compute the $95\%$ CL limits for Higgs observables, both in the full 2HDM and in the SMEFT matched to the 2HDM. In the full 2HDM, we calculate the observables both at LO and at one-loop (NLO). More specifically, we approximate the NLO results \cite{Kanemura:2015mxa, Altenkamp:2017ldc, Kanemura:2017wtm, Altenkamp:2017kxk, Chen:2018shg, Chen:2019pkq} by 
considering the case near the alignment limit (analytic expressions used to obtain our NLO curves, reproduced from Ref.~\cite{Kanemura:2015mxa}, are given in Appendix~\ref{app:zs}). 
We have not considered uncertainties in the full 2HDM calculation due to possible resummation of the  logarithms or uncertainties in the SMEFT due to the truncation of the expansion.
In the SMEFT matched to the 2HDM, we consider three types of contributions: $d_6^{\rm tree}$, $d_8^{\rm tree}$ and $d_6^{\rm loop}$. The first two consider only the operators generated via tree-level matching, such that $d_6^{\rm tree}$ ($d_8^{\rm tree}$) includes only $\mathcal{O}(\xi^1)$ ($\mathcal{O}(\xi^2)$) effects, while $d_6^{\rm loop}$ considers only the contributions generated via one-loop matching at $\mathcal{O}(\xi^1)$. Note that at $\mathcal{O}(\xi^2)$ we include $1/\Lambda^4$ terms arising from the squared amplitude in our calculation of the signal strengths, which compete with the dimension-8 contributions at the same order.

\begin{figure}[htb!]
\includegraphics[width=.49\linewidth]{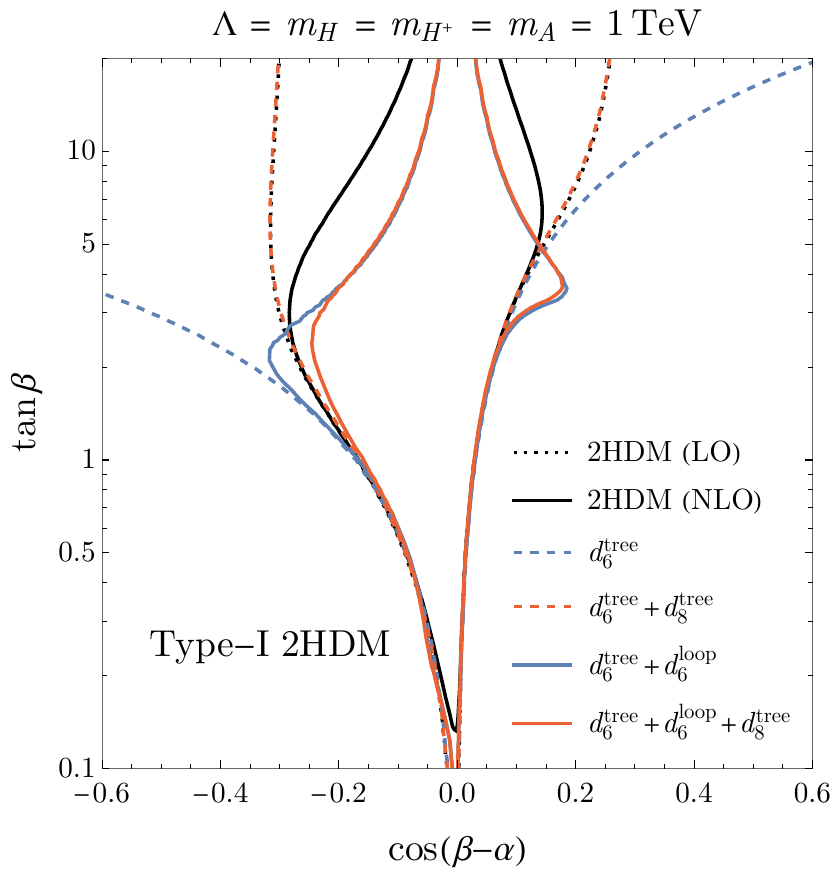}~~
\includegraphics[width=.49\linewidth]{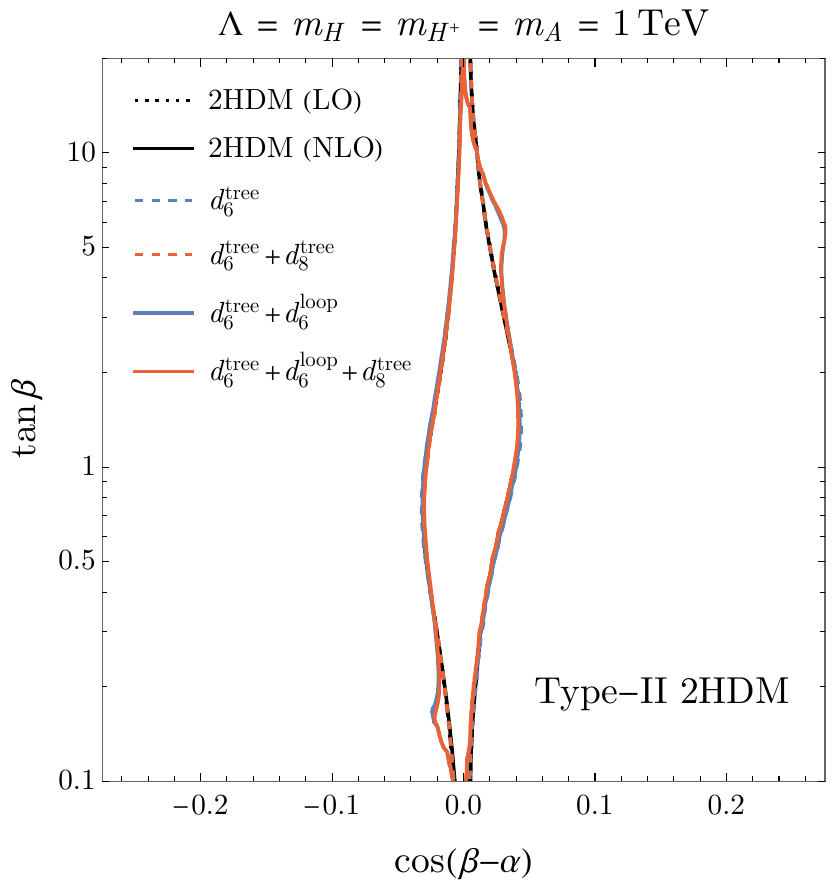}\\[1em]
\includegraphics[width=.49\linewidth]{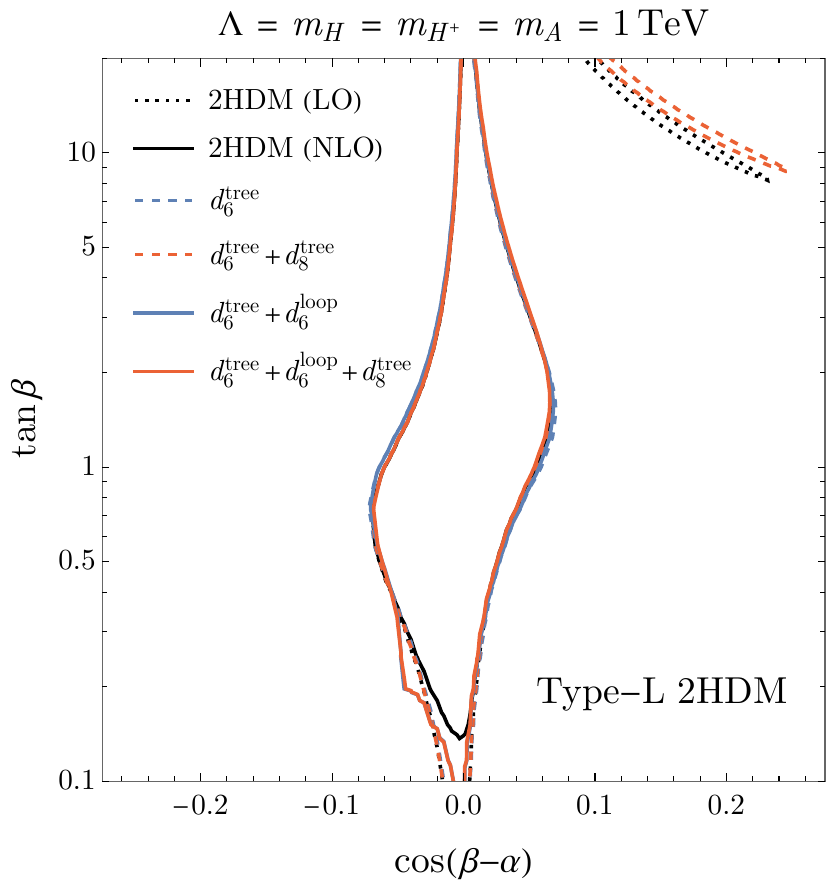}~~
\includegraphics[width=.49\linewidth]{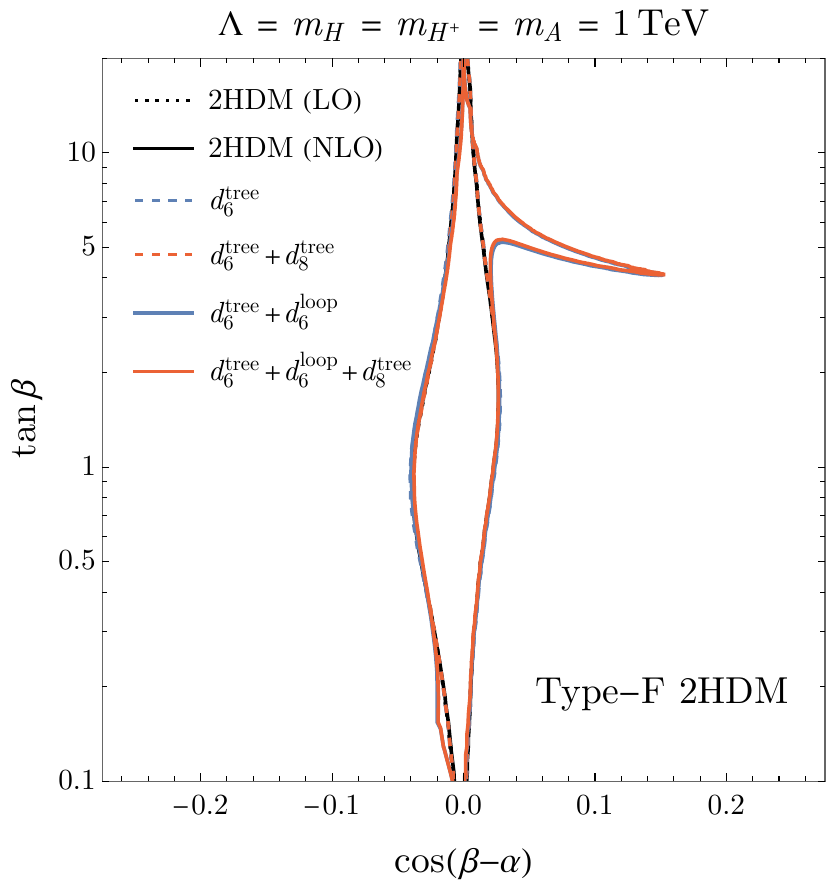}
\caption{
Comparison of the constraints from Higgs precision observables on the 2HDM
with the constraints set by a fit to the SMEFT coefficients.
The dotted and solid black lines show the constraints in the exact model with the Higgs coupling modifiers evaluated at LO and NLO, respectively (see Appendix~\ref{app:zs} for details). The red and blue lines show the SMEFT result with and without the inclusion of the dimension-eight operators ($d_8^{\textrm{tree}}$). The solid red and blue lines include the contributions to the dimension-six operators generated at one-loop in the SMEFT matching ($d_6^{\textrm{loop}}$), while the dashed ones do not. In all four panels, the additional heavy states are assumed to be degenerate and equal to the matching scale, $\Lambda$. Note that the EFT expansion is only formally consistent if $c_{\beta-\alpha}$ is small (i.e. close to the alignment limit, $c_{\beta-\alpha}=0$).
}
\label{fig:higgs-bounds}
\end{figure}

We now turn to the results of Fig.~\ref{fig:higgs-bounds}, considering first Type-I (upper left panel).
Let us start by discussing the results without NLO effects, both in the 2HDM (dotted lines) and in the SMEFT matched to the 2HDM (dashed lines)\fn{We do not show the theoretically allowed regions in these plots, but note that unitarity and perturbativity require that $\cos(\beta-\alpha)$ be close to the alignment limit.} . These results agree with those of Ref.~\cite{ATLAS:2024lyh}. They present slight differences when compared to those of Ref.~\cite{Dawson:2022cmu} (due to the inclusion of new LHC data), but the pattern described above is observed, namely: while the tree-level matching restricted to dimension-6 operators (dashed blue) does not replicate the LO Type-I 2HDM (dotted black) for $\tan \beta > 1$, the inclusion of dimension-8 effects (dashed red) corrects that deficiency.%
\fn{The curve of the tree-level matched to the  dimension-8 operators is essentially on top of the 2HDM LO curve, for the whole range of $\tan \beta$ displayed.}
The inclusion of NLO effects in the full 2HDM fits (solid black) becomes relevant for $\tan \beta \gtrsim 2$, allowing a considerably more restricted range of values of $\cos(\beta-\alpha)$ for larger values of $\tan \beta$ than the LO result. Interestingly, even if restricted to dimension-6 operators, the SMEFT matched to the 2HDM at one-loop (solid blue) captures the essence of that behavior. This effect is discussed in detail below. Note that adding the dimension-8 tree-level matching to the loop matching with dimension-6 operators does not significantly change the latter.%
\fn{NLO effects in the full 2HDM fits introduce a new effect: a lower bound for $\tan \beta \simeq 0.15$. We checked that the SMEFT loop matching also introduces a lower bound, although for lower values of $\tan \beta$ (not visible in the plot).}

As for the three remaining panels of Fig.~\ref{fig:higgs-bounds}, we again confirm what was found in Ref.~\cite{Dawson:2022cmu}: if the region centered around $\cos(\beta-\alpha)\sim 0$
is considered at LO, the tree-level matching with dimension-6 operators is enough to replicate the 2HDM result.
In fact, the tree-level matching with dimension-6 operators (dashed blue) cannot be distinguished from the tree-level matching with dimension-8 operators (dashed red), nor from the LO 2HDM curve (dotted black).
The plots show that the inclusion of loop effects leads to no significant changes either in the full 2HDM results, or in the SMEFT ones.%
\fn{The panels  for models other than the Type-I case show that the loop effects in the SMEFT matching (solid blue and red curves) introduce some kinks not present in the full 2HDM  NLO curves.}
Again as in Ref.~\cite{Dawson:2022cmu}, Type-L is the only type admitting a wrong-sign region,
%(centered around $\cos(\beta-\alpha)\sim 0.15$),
and only when $\mathcal{O}(\xi^2)$ effects are included. With this inclusion, and ignoring NLO effects, the SMEFT (dashed red) reproduces quite well the 2HDM (dotted black). On the other hand, the wrong-sign region vanishes from both the 2HDM and the SMEFT descriptions if loop effects (solid lines) are included.\footnote{In Ref.~\cite{Han:2020lta}, which uses the full one-loop predictions for the Higgs signal strengths, the wrong sign region is allowed in the Type-L and Type-F models (assuming slightly smaller heavy Higgs masses). We attribute this difference to our use of the approximate expressions in \eqref{eq:kappa-1loop}.}

\begin{figure}[htb!]
\centering
\vskip 0.25em
\includegraphics[width=.49\linewidth]{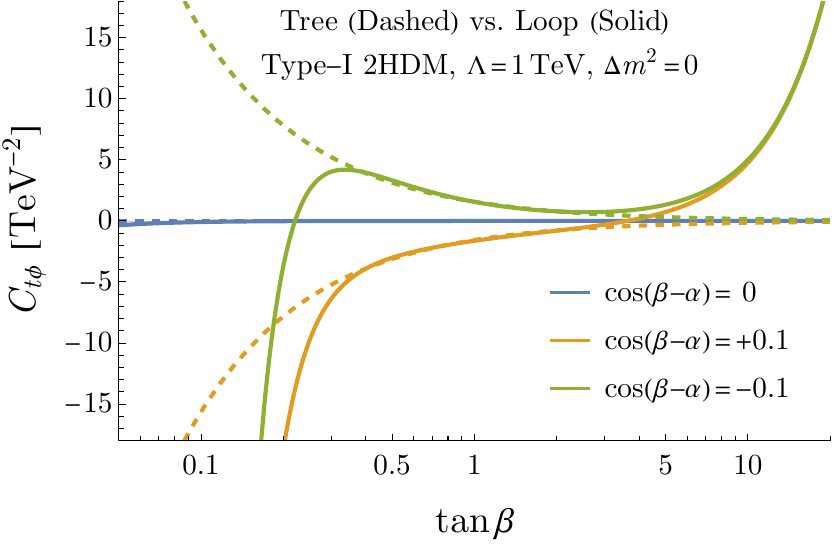}~~
\includegraphics[width=.49\linewidth]{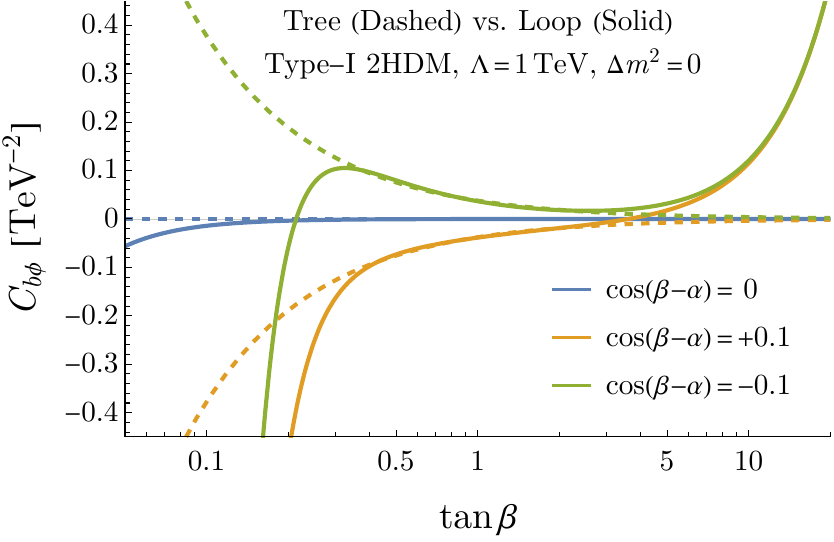}\\[1em]
\includegraphics[width=.49\linewidth]{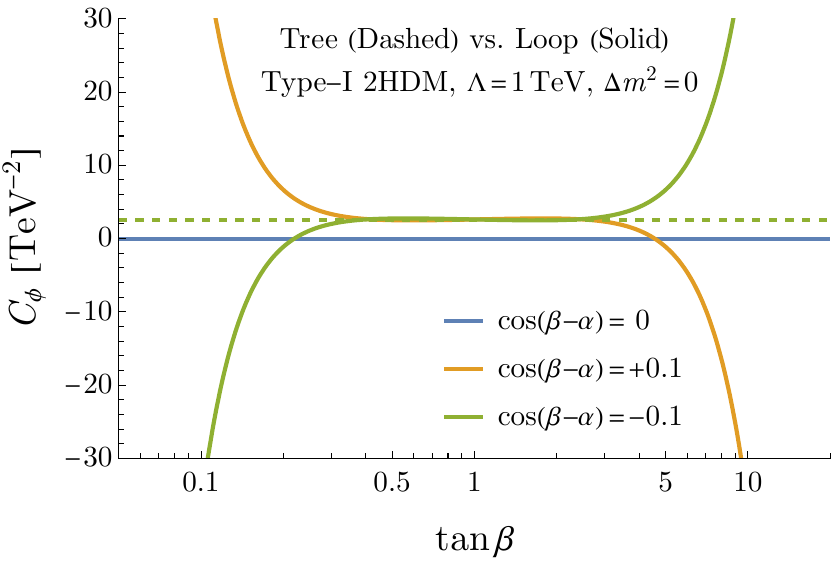}
\caption{
Plots of the WCs $C_{tH}$, $C_{bH}$, and $C_{\phi}$ vs. $\tan\beta$, that are generated in the Type-I 2HDM at tree-level (dashed lines) and one-loop (solid lines), for various values of the alignment parameter, $\cos(\beta-\alpha)$.
}
\label{fig:wc-plots}
\end{figure}

We now discuss the effects observed in the loop matching effects of Type-I. These can be explained by the $C_{f \phi}$ WCs, plotted in the upper half of Fig.~\ref{fig:wc-plots}. In the tree-level matching of Type-I, and as discussed above, they tend to zero at large values of $\tan \beta$. However, in the loop matching and for $\cos(\beta-\alpha) \neq 0$, this behavior is reversed, and the WCs start to grow for $\tan \beta \gtrsim 5$.%
\fn{This effect can be attributed to the parameters $Z_2$ and $Z_7$ of the potential. From Eqs. (\ref{eq:Zs}), these are the only ones that introduce a dependence on $\tan \beta$. Those parameters are not generated in the tree-level matching (up to dimension-8), as they multiply at least three powers of the heavy doublet. This is not the case in the loop matching, where the parameters do contribute, and thus introduce dependences on $\tan \beta$ that turn out to be highly relevant for large values of that parameter.}
Note that, even though the sign of $C_{f \phi}$ is the same as that of $\cos(\beta-\alpha)$ for $\tan \beta \lesssim 5$, it always becomes positive for $\tan \beta \gtrsim 5$.
In Fig.~\ref{fig:wc-plots}, we also show $C_\phi$, although its contributions to Higgs observables are not included in Fig.~\ref{fig:higgs-bounds}. Like $C_{u \phi}$ and $C_{d \phi}$, it acquires large values (in modulus) for large $\tan \beta$ with loop matching only. 

\begin{figure}[htb!]
\includegraphics[width=.49\linewidth]{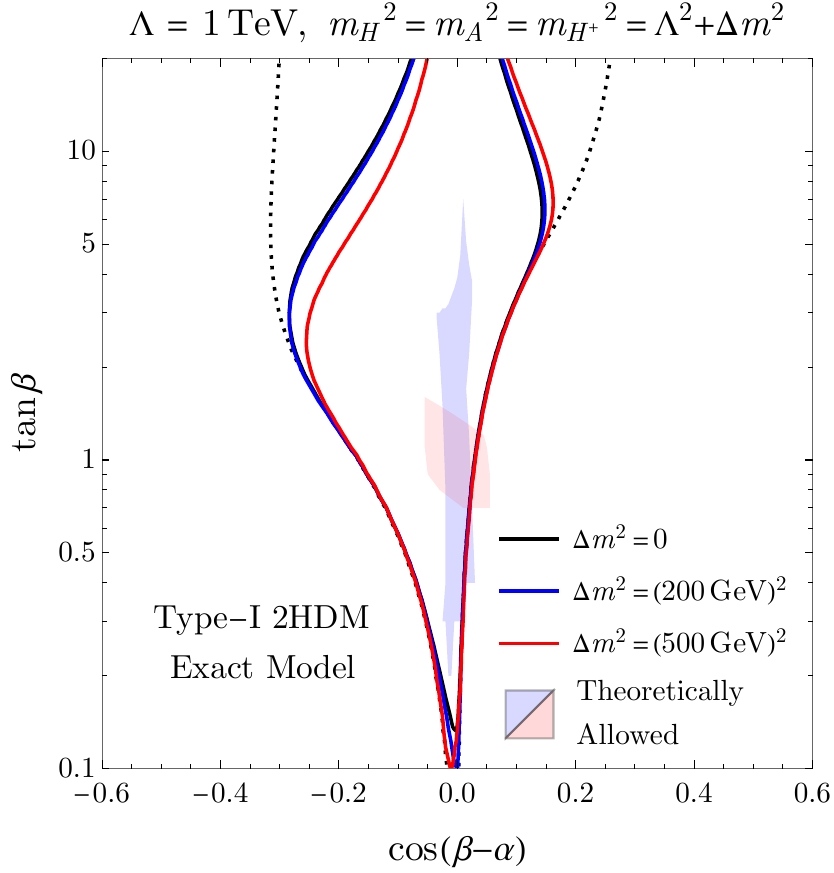}~~
\includegraphics[width=.49\linewidth]{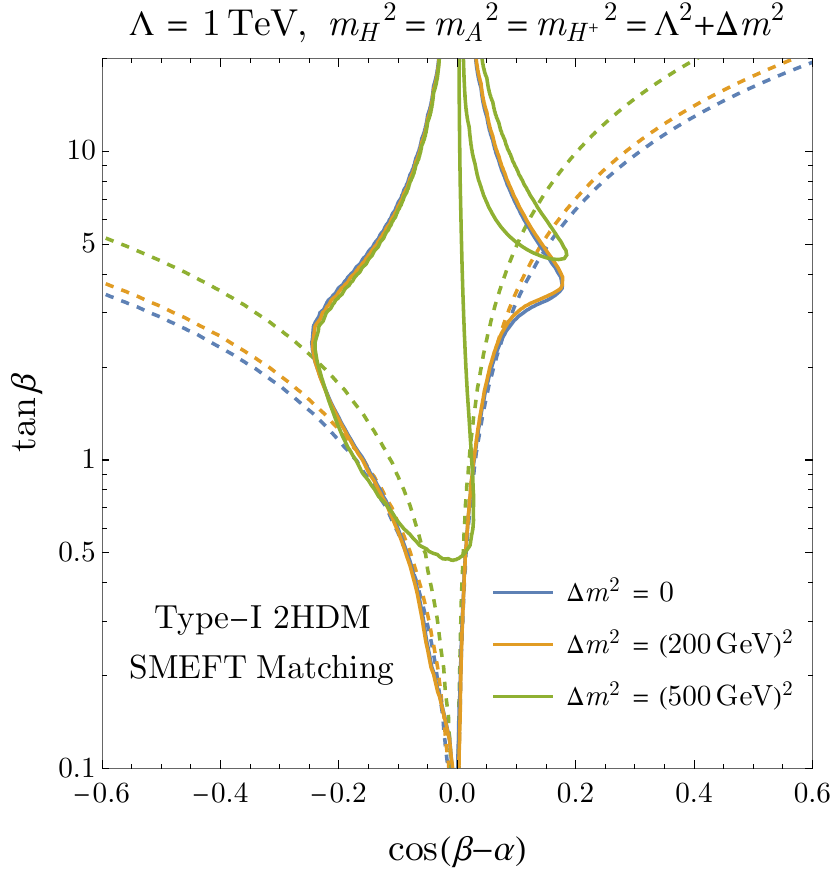}\\[1em]
\includegraphics[width=.49\linewidth]{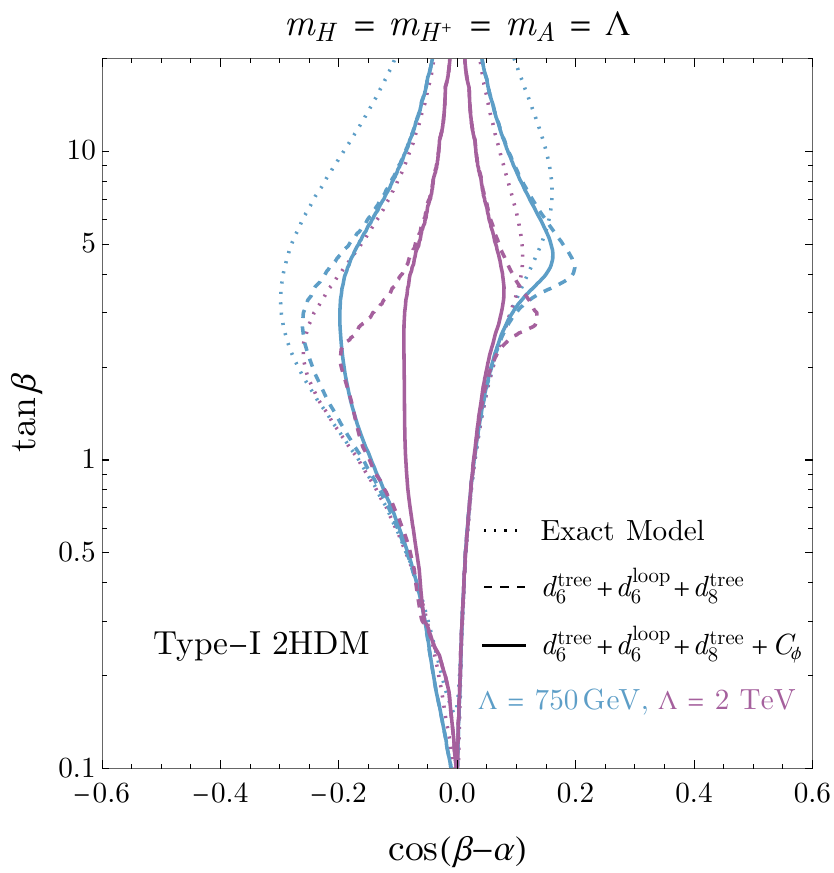}~~
\includegraphics[width=.49\linewidth]{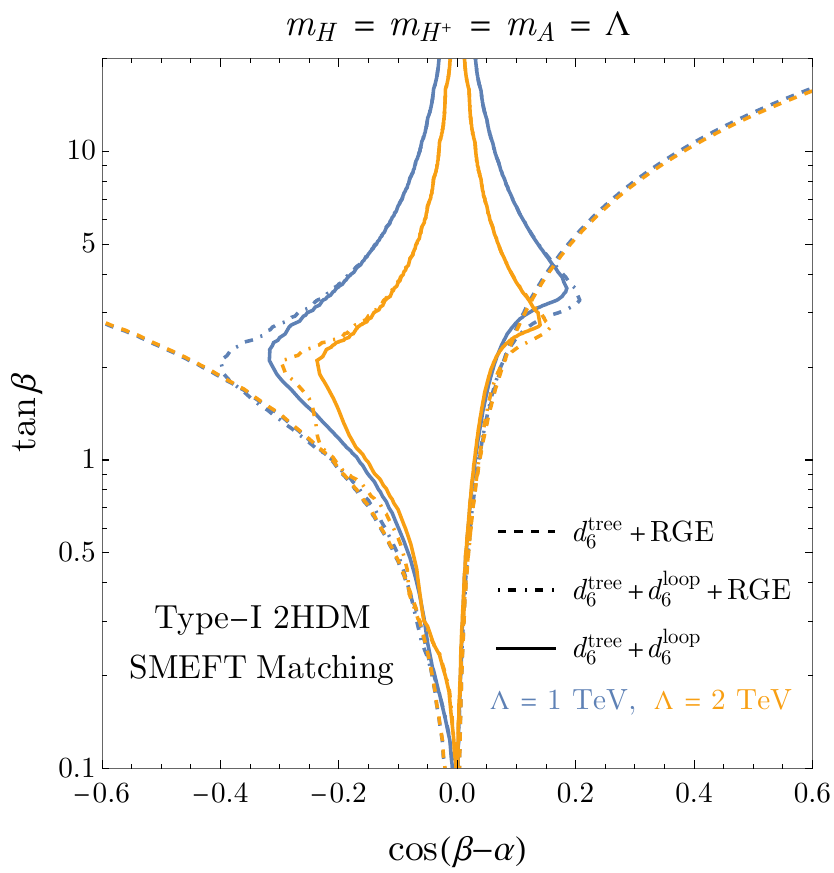}
\caption{
Additional plots showing the constraints from Higgs measurements at the LHC on the fits to the Type-I 2HDM. The upper left panel shows the constraints on the exact model (using the NLO predictions, with the heavy Higgs states separated from the matching scale by various values of $\Delta m^2$) and compares these constraints to the theoretically allowed regions described in the text. The LO prediction for the exact model (which does not depend on the mass splitting) is shown as a dashed black curve for comparison.
In the upper right panel, we show the same curves,
now evaluated using the SMEFT matching at dimension-6 at tree-level (dashed) and one-loop (solid).
In the lower left panel, we show the effects of including indirect information from single Higgs production on the Higgs self coupling ($C_\phi$) for different values of the matching scale. The lower right panel illustrates the effects of including RGE of the WCs generated at tree-level in the 2HDM, and compares these to the effects of the full contributions generated at one-loop. Note that the EFT expansion is only formally consistent if $c_{\beta-\alpha}$ is small (i.e. close to the alignment limit, $c_{\beta-\alpha}=0$).
}
\label{fig:type1-extras}
\end{figure}

Given the relevance of Type-I for understanding the accuracy of the SMEFT matching to the 2HDM, we explore different features of the Type-I model in Fig.~\ref{fig:type1-extras}. We define $\Delta m^2 \equiv \Delta m_H^2 = \Delta m_A^2 = \Delta m_{H^{\pm}}^2$.
In the upper left panel, we consider just the full 2HDM results (not the SMEFT), and investigate both regions of parameter space allowed by theoretical constraints, as well as the dependence on $\Delta m^2$. It is clear that the allowed regions are extremely constrained, and forced to be very close to both $\cos(\beta-\alpha)=0$ and $\tan \beta=1$.%
\fn{For $\Delta m^2=0$, the allowed region is so small that we do not show it.}
On the other hand, the 2HDM curves do not depend significantly on $\Delta m^2$. The upper right panel considers again the dependence on $\Delta m^2$, but now using SMEFT matching to the 2HDM. With tree-level matching, larger values of $\Delta m^2$ push the curves upwards in $\tan \beta$. With loop matching, there is little difference between $\Delta m^2=0$ and $\Delta m^2=(\rm{200 \, GeV})^2$, but a significant difference for $\Delta m^2=(\rm{500 \, GeV})^2$. It is clear that the dimension-6 tree level curves (dotted) cannot reproduce the full model results, since they do not include information from the Higgs couplings to vector bosons. When the loop effects are included (solid), there is a coupling of the Higgs to vector bosons which grows with $\tan\beta$ and so the generic shape of the full 2HDM is recovered by the SMEFT fit.  It is important to note that loop effects in both the full 2HDM and the SMEFT are necessary for this agreement.  In the region allowed by theoretical constraints in the full model, $\cos(\beta-\alpha)\sim 0$, there is excellent agreement between the full 2HDM and the SMEFT fit.

Still in Fig.~\ref{fig:type1-extras}, the lower left panel explores both a dependence on the scale $\Lambda$, as well as the inclusion of $C_\phi$ effects (which allow us to indirectly determine the
Higgs self-interactions from single Higgs production\cite{Degrassi:2016wml,Degrassi:2021uik}).%
\fn{For details on the inclusion of $C_\phi$ effects, see Ref.~\cite{Dawson:2022cmu}.}
As expected, larger values of $\Lambda$ push the results closer to the alignment limit. The inclusion of $C_\phi$ effects restricts the results even further.
Finally, the lower right panel of Fig.~\ref{fig:type1-extras} investigates the role of renormalization group evolving the dimension-6 operators between the scales $\Lambda$ and $m_Z$. The plot shows that, without loop matching, the RGEs play a very little role (when compared to the dashed blue curve of the upper left panel of Fig.~\ref{fig:higgs-bounds}), and have a very small dependence on the scale. On the other hand, when loop matching effects are included, the effects of including the RGEs are similar to those of changing $\Lambda$ by a TeV.

\section{Conclusions}
\label{sec:conclusions}

We have performed for the first time the complete one-loop matching of the 2HDM to the SMEFT with dimension-6 operators. This generates numerous operators not present with the tree-level matching. We derived our result using the  software \texttt{Matchmakereft} and \texttt{Matchete}. We obtained agreement in our results between them, which provides a non-trivial check of both software packages. Auxiliary files accompanying this paper include detailed results of the matching.

We demonstrated how the notion of decoupling allowed us to perform a consistent expansion in terms of more physical parameters. This led us to compare the one-loop matching with dimension-6 operators to the tree-level matching with dimension-8 operators derived in Ref.~\cite{Dawson:2022cmu}. We did this by performing fits to both EWPOs and to Higgs observables. In the case of EWPOs, while none of the dimension-8 operators generated at tree-level contribute, the dimension-6 loop-generated operators do, thus clearly showing the need for  the latter type of matching. In the fits to Higgs observables, we studied the four types of 2HDM and showed that, for Type-II, Type-L and Type-F, the inclusion of higher order effects is not significant, as both the full model 2HDM description and the SMEFT matching to the 2HDM are barely affected by it.

In Type-I, by contrast, the 2HDM results for larger $\tan \beta$ evaluated at NLO are quite different from those evaluated at LO. We showed that the 
SMEFT matched with loop-generated dimension-6 operators captures most of the physics, and so provides an overall adequate replication of the full model results. We identified the Warsaw basis operators $C_{u \phi}$ and $C_{d \phi}$ as mainly responsible for this phenomenon, as their values at large $\tan \beta$ become quite large with one-loop matching only. We investigated how the results depend on the heavy scale $\Lambda$, on the difference between the heavy masses and  $\Lambda$, and on the RGE effects. In all cases, we found significant effects in the region of larger $\tan \beta$.

We note that the difference in the full model between LO and NLO results shows that, in the SMEFT approach, the results obtained with one-loop matching and dimension-6 operators should not be directly compared to those obtained with tree-level matching and dimension-8 operators. Current LHC experimental fits to the 2HDM are performed using tree level predictions, and should be updated to include the NLO 2HDM results. Our analysis also shows that, when one requires a separation of scales that allows an EFT description, the 2HDM has a very small region of parameter space allowed by theoretical constraints. In particular, the regions where the non-trivial SMEFT truncations (both at tree-level with dimension-8 operators and at one-loop with dimension-6 operators) become relevant are excluded. On the other hand, matching at one-loop level has been performed in a singlet extension of the SM, only to conclude that it is not relevant in such a simple extension. Both conclusions lead us to suggest an investigation of loop matching in richer UV models, where such matching can play a more decisive role, would be of interest.

Furthermore, although we presented the complete one-loop matching between the 2HDM and the SMEFT with dimension-6 operators, our work represents only a first step towards a complete NLO analysis. Indeed, our analysis includes only approximate expressions of the full NLO 2HDM, and does not contain one-loop correction to the observables computed in the SMEFT. These shortcomings motivate future studies that might ascertain the importance of those corrections.
The auxiliary files for this project can be accessed at the following URL: \url{https://github.com/BDFH-2024/BDFH}.
\section*{Ackowledgments}
We thank Pier Paolo Giardino for discussions. S.D.B. thanks  Javier Fuentes-Mart\'in and Jos\'e Santiago for helping with the SMEFT matching results computation and cross-verification. S.D.B. and D.F. are grateful to the Mainz Institute for Theoretical Physics (MITP) of the Cluster of Excellence PRISMA+ (Project ID 39083149) for its hospitality and support.
S.D.B is supported by SRA (Spain) under Grant No. PID2019-106087GB-C21 (10.13039/501100011033) and  PID2021-128396NB- 100/AEI/10.13039/501100011033; by the Junta de Andalucía (Spain) under Grants No. FQM-101 and P21\_00199. %
S.D. and D.F. are supported by the U.S. Department of Energy under Grant Contract No. DE-SC0012704. S.H. is supported by the DOE grant DE-SC0013607.

\clearpage
\appendix

\renewcommand{\thesection}{\Alph{section}}
\renewcommand{\theequation}{\thesection.\arabic{equation}}
\setcounter{equation}{0}

\section{More details on the 2HDM}
\label{app:zs}

From the minimization of the potential,
\begin{equation}
\label{eq:theYs}
Y_1 = - \dfrac{Z_1}{2} v^2,
\qquad
Y_3 = - \dfrac{Z_6}{2} v^2 \, .
\end{equation}
Expressing the $Z_i$ parameters of Eq. (\ref{eq:potential}) in terms of the parameters of \eqref{eq:indep-real}, 
\begin{eqnarray}
\label{eq:Zs}
Z_1 &=& \dfrac{\sba^2 m_{h}^2 + \cba^2 \, m_{H}^2}{v^2}, \nonumber \\[0.25em]
Z_2 &=&
\dfrac{1}{2 \, v^2 \, t_{\beta}^3} \bigg[
\cba^2 \, t_{\beta} \left(3 t_{\beta}^4 - 8 t_{\beta}^2 + 3\right)
 \left(m_{h}^2 - m_{H}^2\right) 
+ \sba \, \cba \big(t_{\beta}^6 - 7 t_{\beta}^4 + 7 t_{\beta}^2 -1\big) \big(m_{h}^2-m_{H}^2\big), \nonumber\\
&& \hspace{13mm} 
- m_{h}^2 \left(t_{\beta}^5 - 4 t_{\beta}^3 + t_{\beta}\right) + 2 t_{\beta} \left(t_{\beta}^2 - 1\right)^2 \left(m_{H}^2 - Y_2\right)
\bigg], \nonumber \\[0.25em]
Z_3 &=& \dfrac{2}{v^2} \left(m_{H^{+}}^2 - Y_2\right), \nonumber \\[0.25em]
Z_4 &=& \dfrac{\cba^2 \left(m_{h}^2 - m_{H}^2\right) + m_A^2 + m_{H}^2 - 2 \, m_{H^{+}}^2}{v^2}, \nonumber \\[0.25em]
Z_5 &=& \dfrac{\cba^2 \left(m_{h}^2 - m_{H}^2\right) - m_A^2 + m_{H}^2}{v^2}, \nonumber \\[0.25em]
Z_6 &=& \dfrac{\cba \, \sba \, \left(m_{h}^2 - m_{H}^2\right)}{v^2}, \nonumber \\[0.25em]
Z_7 &=& \dfrac{1}{2 \, v^2 \, t_{\beta}^2} \bigg[
-3 \cba^2 t_{\beta} \left(t_{\beta}^2-1\right)
 \left(m_{h}^2-m_{H}^2\right)- \sba
 \cba \left(t_{\beta}^4-4 t_{\beta}^2+1\right)
 \left(m_{h}^2-m_{H}^2\right), \nonumber \\
 && \hspace{13mm} +t_{\beta}
 \left(t_{\beta}^2-1\right) \left(m_{h}^2-2 m_{H}^2+2
 Y_2\right)
\bigg].
\label{eq:zmass}
\end{eqnarray}
The effects of the 2HDM on measurements of the $125\,\textrm{GeV}$ Higgs boson production and decay processes can be conveniently parameterized by the so-called ``$\kappa$-framework'', where the Higgs couplings to other SM particles are rescaled by Higgs coupling modifiers, $\kappa_f$, $\kappa_V$ ($V = W, Z$), such that the SM prediction is recovered when $\kappa = 1$.

The one-loop (NLO) corrections to the scaling of the Higgs couplings in the 2HDM can be approximated by including
contributions which grow with $m_t$ and the heavy scalar masses.  Working near the decoupling limit, we have~\cite{Kanemura:2015mxa}
\begin{eqnarray}
\label{eq:kappa-1loop}
\kappa_V & = & \sba
- \frac{1}{6} \frac{1}{16\pi^2} 
\Bigg[ \frac{m_H^2}{v^2} \Big(1 - \frac{M^2}{m_H^2}\Big) + 2 \frac{m_{H^+}^2}{v^2}\Big(1 - \frac{M^2}{m_{H^+}^2}\Big) + \frac{m_A^2}{v^2} \Big(1 - \frac{M^2}{m_A^2} \Big) \Bigg],
\nonumber \\[0.5em]
\kappa_t & = & \kappa_V + \frac{\eta_u \cba}{\tan\beta}
- \frac{1}{3 \tan^2\beta} \frac{1}{16\pi^2} \Bigg[ \frac{\eta_u^2 m_t^2}{v^2} \Big( \frac{m_t^2}{m_H^2} + \frac{m_t^2}{m_{H^+}^2} + \frac{m_t^2}{m_A^2} \Big) + \eta_d^2 \frac{m_b^2 m_t^2}{v^2 m_{H^+}^2} \Bigg] ,
\nonumber \\[0.5em]
\kappa_b & = & \kappa_V + \frac{\eta_d \cba}{\tan\beta} 
- \frac{\eta_u \eta_d}{\tan^2\beta} \frac{1}{16\pi^2} \frac{4 m_t^2}{v^2} \Bigg[ 1 - \frac{M^2}{m_{H^+}^2} + \frac{m_t^2}{m_{H^+}^2}\Big(1 + \log\frac{m_t^2}{m_{H^+}^2} \Big) \Bigg] \nonumber\\
&& \quad - \frac{1}{3} \frac{\eta_d^2 m_b^2}{v^2 \tan^2\beta} \frac{1}{16\pi^2} \bigg( \frac{m_b^2}{m_H^2} + \frac{m_b^2}{m_{H^+}^2} + \frac{m_b^2}{m_A^2} \bigg) ,
\nonumber \\[0.5em]
\kappa_\tau & = & \kappa_V + \frac{\eta_e \cba}{\tan\beta} ,
\end{eqnarray}
where
\be
M^2 = Y_2 - Y_1 - 2 Y_3 \tan 2\beta,
\ee
is the scale that describes the soft-breaking of the $\mathbb{Z}_2$ symmetry.
We emphasize again that these are not the complete expressions for the loop corrections to the Higgs couplings, only the leading terms in an expansion in $x \equiv \pi/2 - (\beta - \alpha) \ll 1$. In particular, we keep only the $x$-independent one-loop corrections. See Ref.~\cite{Kanemura:2015mxa} for more details.

\vskip -1em

\section{One-loop matching results}
\label{app:loop matching}

In this appendix, we present the matching expressions for the SMEFT coefficients of \eqref{eq:L6-loop}. As discussed in Section~\ref{sec:decoupling}, because these coefficients are loop generated, we expand them only to $\mathcal{O}(\xi^1)$.
%Although in general we do not write explicitly the operators involving leptons, we do write their coefficients explicitly; for whereas the operators are trivially obtained from those involving down-type quarks, the same does not happen in general for the coefficients. 
We recall (\eqref{eq_t-vs-l}) that, for operators that are not generated at tree-level, we write the coefficients as $C_i$ (without superscript $[l]$).  For operators that are generated at tree-level, $C_\phi$ and $C_{f\phi}$, we give the one-loop contributions, $C_\phi^{[l]}$ and $C_{f\phi}^{[l]}$, here, which must be added to the dimension-6 tree-level contributions,$C_\phi^{[t]}$ and $C_{f\phi}^{[t]}$, of \eqref{eq:WC-dim6-tree} to obtain the full results. As always, the generation indices are suppressed.

The results are:
\bs
\bea
\dfrac{C_{\phi W}}{\Lambda^2} &=& \dfrac{m_W^2 \, G_F^2}{96 \, \pi^2 \, \Lambda^2} \left( \Delta m_{A}^2 + \Delta m_{H}^2 + 2 \, \Delta m_{H^{+}}^2 \right), \\[0.5em]
\dfrac{C_{\phi B}}{\Lambda^2} &=& \dfrac{G_F^2}{96 \, \pi^2 \, \Lambda^2} (m_Z^2 - m_W^2) \left( \Delta m_{A}^2 + \Delta m_{H}^2 + 2 \, \Delta m_{H^{+}}^2 \right), \\[0.5em]
%
%
%
%\dfrac{C_{\phi d}}{\Lambda^2} &=& \dfrac{G_F^2}{96 \, \pi^2 \, \Lambda^2} \left( m_Z^2 - m_W^2 \right) \left( \Delta m_{A}^2 + \Delta m_{H}^2 + 2 \, \Delta m_{H^{+}}^2 \right) \\
%
%
%
\dfrac{C_{\phi WB}}{\Lambda^2} &=& \dfrac{G_F^2 {m_W} \, {m_Z}}{48 \,\pi^2 \, \Lambda^2} \sqrt{1 - \dfrac{m_W^2}{m_Z^2}} \left( \Delta m_{A}^2 + \Delta m_{H}^2 - 2 \, \Delta m_{H^{+}}^2 \right), \\[0.5em]
\dfrac{C_\phi^{[l]}}{\Lambda^2} &=& -\dfrac{(\sqrt{2} G_F)^3 \, \cot^3 \beta}{480 \, \pi^2 \, \Lambda^2} \Bigg\{ -\dfrac{45}{2} \, \cba^3 \, \Lambda^6 \left( \cos(2 \, {\beta}) + \cos(6 \, {\beta}) \right) \sec^6 \beta \nonumber \\
&& + 2 \, \bigg( 10 \left( \Delta m_{A}^6 + \Delta m_{H}^6 + 2 \, \Delta m_{H^{+}}^6 \right) - 5 \, \Big[ \Delta m_{A}^4 + \Delta m_{H}^4 - 2 \left( \Delta m_{A}^2 + \Delta m_{H}^2 \right) \, \Delta m_{H^{+}}^2 \nonumber \\
&& + 2 \, \Delta m_{H^{+}}^4 \Big] \, m_h^2 + 2 \, m_h^2 \, m_W^4 \bigg) \tan^3 \beta 
+ 15 \, \cba^2 \, \Lambda^4 \, \tan\beta \Big[ m_h^2 \left( -5 + 2 \, \cos(4 \, {\beta}) \right) \, \sec^4 \beta 
\nonumber \\
&& + 6 \, \Delta m_{H}^2 \left( -1 + \sec^2 \beta + \sec^4 \beta \right) - 2 \left( \Delta m_{A}^2 + 2 \, \Delta m_{H^{+}}^2 \right) \, \tan^2 \beta \Big]  \nonumber \\
&& - 90 \, \cba \, \Lambda^2 \left( -2 \, \Delta m_{H}^2 + m_h^2 \right) ^2 \, \tan^2 \beta \left( \tan^2({\beta})-1\right) \Bigg\}, \\[0.5em]
\dfrac{C_{\phi \Box}}{\Lambda^2} &=& -\dfrac{G_F^2}{480 \, \pi^2 \, \Lambda^2} \, \Big[ -5 \left( \Delta m_{A}^2 - \Delta m_{H}^2 \right) ^2 + 20 \left( \Delta m_{A}^2 + \Delta m_{H}^2 \right) \, \Delta m_{H^{+}}^2 \nonumber \\
&& \hspace{24mm} + 2 \left( 45 \, \cba^2 \, \Lambda^4 + 4 \, m_W^4 - 2 \, m_W^2 \, m_Z^2 + m_Z^4 \right) \Big], \\[0.5em]
\dfrac{C_{\phi D}}{\Lambda^2} &=& -\dfrac{G_F^2}{60 \, \pi^2 \, \Lambda^2} \, \Big[ -5 \, \Delta m_{H}^2 \, \Delta m_{H^{+}}^2 + 5 \, \Delta m_{H^{+}}^4 + 5 \, \Delta m_{A}^2 \left( \Delta m_{H}^2 - \Delta m_{H^{+}}^2 \right) \nonumber \\
&& \hspace{24mm} + m_W^4 - 2 \, m_W^2 \, m_Z^2 + m_Z^4 \Big], \\[0.5em]
\dfrac{C_{u\phi}^{[l]}}{\Lambda^2} &=& \dfrac{{m_u} \, (\sqrt{2} G_F)^{5/2}}{\sqrt{2} \, 480 \, \pi^2 \, \Lambda^2}
 \Bigg\{
10 \, \Big[ \Delta m_{A}^4 + (1-36 \nnu) \, \Delta m_{H}^4 - 2 \, \Delta m_{A}^2 \, \Delta m_{H^{+}}^2 + 2 \, \Delta m_{H^{+}}^4 
- 9 \, \nnu \, m_h^4
\nonumber \\
&& 
- (30 + 63 \nnu) \, \cba^2 \, \Lambda^4  - 2 \, \Delta m_{H}^2 \left( \Delta m_{H^{+}}^2 - 18 \nnu \, m_h^2 \right) \Big] 
- 30 \, \cba \cot\beta \, \Lambda^2 \, \Big( 2 \, \nnu \, \Delta m_{A}^2\nonumber \\ 
&& 
- 6 (1+ \nnu) \, \Delta m_{H}^2 + 4 \, \nnu \, \Delta m_{H^{+}}^2 -  \, m_d^2 ( 3 \nnd + 4 \nnu)  + 3 \, m_h^2 + 13 \, \nnu \, m_u^2 \Big) - 4 \, m_W^4  \nonumber \\
&& + 10 \, \cot^2 \beta \, \Big[ 6 \big( 6 \, \nnu \, \Delta m_{H}^4 + \frac{3}{2} \, (1+7 \nnu) \, \cba^2 \, \Lambda^4+ \nnd (4 \nnu - \nnd) \, \Delta m_{H^{+}}^2 \, m_d^2  \nonumber \\
&& - \nnd ( \nnd + 4 \nnu) \, m_d^4 \big)
+ 9 \left( -4 \, \nnu  \, \Delta m_{H}^2 + \nnu \nnd  m_d^2 \right) \, m_h^2 + 9 \nnu \, m_h^4 \nonumber \\
&& - 2 \, \nnu^2 \left( -3 \, \Delta m_{A}^2 + 9 \, \Delta m_{H}^2 + 3 \, \Delta m_{H^{+}}^2  \right) \, m_u^2
- 2 (6 \nnd \nnu + 3 \nnu^2 - \nnd^2) m_d^2 m_u^2 - 6 \nnu^2 \, m_u^4 \Big] \nonumber \\
&& + 15 \nnu \, \cba \, \Lambda^2 \left( -7 \nnd^2 \, m_d^2 + 3 \nnu^2 \, m_u^2 \right) \cot^3 \beta  
- 90 \nnu \, \cba^2 \, \Lambda^4 \, \cot^4 \beta \nonumber \\
&& + 90 \, \cba  \Lambda^2  \tan({\beta})  \Big[ -2  \Delta m_{H}^2 + m_h^2 + (1+\nnu)  \cba  \Lambda^2  \tan({\beta}) \Big] \Bigg\}, \\[0.5em]
\dfrac{C_{d\phi}^{[l]}}{\Lambda^2} &=& \dfrac{{m_d} \, (\sqrt{2} G_F)^{5/2}}{\sqrt{2} \, 480 \, \pi^2 \, \Lambda^2 }
\bigg\{
10 \, \Big[ \Delta m_{A}^4 +(1- 36 \nnd) \, \Delta m_{H}^4 - 2 \, \Delta m_{A}^2 \, \Delta m_{H^{+}}^2 + 2 \, \Delta m_{H^{+}}^4  \nonumber \\
&& - (30+63 \nnd) \, \cba^2 \, \Lambda^4 - 9 \nnd \, m_h^4 - 2 \, \Delta m_{H}^2 \left( \Delta m_{H^{+}}^2 - 18 \, \nnd \, m_h^2 \right) \Big]  - 4 \, m_W^4 \nonumber \\
&&
- 30 \, \cba\, \Lambda^2 \, \cot\beta \, \Big( 2 \nnd \, \Delta m_{A}^2  
- 6(1+\nnd) \, \Delta m_{H}^2 + 4 \nnd \, \Delta m_{H^{+}}^2 + 13 \nnd \, m_d^2 + 3 \, m_h^2 \nonumber \\
&& - (4 \nnd + 3 \nnu) \, m_u^2 \Big)  
+ 10  \, \cot^2 \beta \, \Big[ 36\nnd \, \Delta m_{H}^4 + 9(1+7 \nnd) \, \cba^2 \, \Lambda^4 + 9 \nnd \, m_h^4 \nonumber \\
&& - 6 \nnd^2 \, m_d^2 \left( -\Delta m_{A}^2 + \Delta m_{H^{+}}^2 + m_d^2 \right) 
- 18 \nnd \, \Delta m_{H}^2 \left( m_d^2 \nnd + 2 \, m_h^2  \right) - 6(\nnu^2 +4\nnd \nnu) \, m_u^4 \nonumber \\
&& 
+ \big( 6 \nnu (4 \nnd - \nnu) \, \Delta m_{H^{+}}^2 + 2 (\nnu^2 - 6 \nnu \nnd - 3 \nnd^2) \, m_d^2 + 9 \nnd \nnu \, m_h^2 \big) \, m_u^2 \Big] \nonumber\\
&& + 15 \nnd \, \cba \, \Lambda^2 \left( 3 \nnd^2 \, m_d^2 - 7 \nnu^2 \, m_u^2 \right) \, \cot^3 \beta 
- 90 \nnd \, \cba^2 \, \Lambda^4 \, \cot^4 \beta \nonumber \\
&& + 90 \, \cba \, \Lambda^2 \, \tan\beta \left( -2 \, \Delta m_{H}^2 + m_h^2 + (1+\nnd) \, \cba \, \Lambda^2 \, \tan({\beta}) \right) \bigg\}, \\[0.5em]
\dfrac{C_{e\phi}^{[l]}}{\Lambda^2} &=& 
\dfrac{{m_e} \, (\sqrt{2} G_F)^{5/2}}{\sqrt{2} \, 240 \, \pi^2 \, \Lambda^2 }
\bigg\{ 5 \Delta m_A^4 - 10 \Delta m_A^2 \Delta m_{H^+}^2 + 10 \Delta m_{H^+}^4 - 2 m_W^4 \nonumber \\
&& - 15 \, \cba \, \Lambda^2 \, \cot \beta \Big( 6 \Delta m_{H}^2 (-1 + \tan^2 \beta - \nnl ) - 3 m_h^2 (-1 + \tan^2 \beta)  \nonumber\\[0.5em]
&& + 2 \nnl (\Delta m_A^2 + 2 \Delta m_{H^{+}}^2) \Big)
- 10 \Delta m_H^2 \Big( \Delta m_{H^+}^2 - 18 m_h^2 (-1 + \tan^2 \beta) \nnl \cot^2 \beta  \Big) \nonumber\\
&&
+ 15 \cba^2 \Lambda^4 \Big(- 3 \nnl \cot^4 \beta + 3 \tan^2 \beta (1 + \nnl ) + 3 (1+7\nnl) \cot^2 \beta \Big. 
\Big.- (10+21\nnl)\Big)  \nonumber\\
&& + 45 m_h^4 \nnl \cot^2 \beta - 45 m_h^4 \nnl 
+ 5 \Delta m_H^4 \Big(  (1 - 36 \nnl ) + 36 \nnl \cot^2\beta \Big) \bigg\}, \\[0.5em]
%
%
%
%\dfrac{C_{\phi u}}{\Lambda^2} &=& - \dfrac{G_F^2}{90 \, \pi^2 \, \Lambda^2} \,   \left( m_W^2 - m_Z^2 \right) ^2  \\
%
%
%
\dfrac{C_{\phi u}}{\Lambda^2} &=& \dfrac{G_F^2}{1080 \, \pi^2 \, \Lambda^2} \, \Big[ -12 \left( m_W^2 - m_Z^2 \right) ^2 + \left\lbrace 5 \, \nnu^2 \, m_u^2 \left( - 63 \, m_u^2 + 10 \, m_W^2 - 10 \, m_Z^2 \right) \right. \nonumber \\
&& \hspace{24mm} \left. - 45 \, m_u^2 m_d^2 \left( \nnd^2 - 6 \nnd \nnu - 6 \nnu^2 \right) \right\rbrace \, \cot^2 \beta \Big], \\[0.5em]
%
%
%
%\dfrac{C_{\phi d}}{\Lambda^2} &=&  \dfrac{G_F^2}{180 \, \pi^2 \, \Lambda^2} \,   \left( m_W^2 - m_Z^2 \right) ^2  \\
%
%
%
\dfrac{C_{\phi d}}{\Lambda^2} &=& \dfrac{G_F^2}{1080 \, \pi^2 \, \Lambda^2} \, \Big[ 6 \left( m_W^2 - m_Z^2 \right) ^2 + \left\lbrace 5 \, \nnd^2 \, m_d^2 \left( 63 \, m_d^2  + 22 \, m_W^2 - 22 \, m_Z^2 \right) \right. \nonumber \\
&& \hspace{24mm} \left. + 45 \, m_d^2 \, m_u^2 \left( -6 \nnd^2 - 6 \nnu \nnd + \nnu^2 \right) \right\rbrace \, \cot^2 \beta \Big], \\ [0.5em]
\dfrac{C_{\phi e}}{\Lambda^2} &=& \dfrac{G_F^2}{360 \, \pi^2 \, \Lambda^2} \left( 6 \left( m_W^2 - m_Z^2 \right) ^2 + 35 \, \nnl^2 \, m_{e}^2 \left( 3 \, m_{e}^2 - 2 \, m_W^2 + 2 \, m_Z^2 \right) \, \cot^2 \beta \right), \\[0.5em]
\dfrac{C_{\phi q}^{(1)}}{\Lambda^2} &=& -\dfrac{G_F^2}{1080 \, \pi^2 \, \Lambda^2} \bigg\{ 3 \left( m_W^2 - m_Z^2 \right)^2 + 5 \cot^2 \beta \Big[ 63 \, \nnd^2 \, m_d^4  \nonumber \\
&& \hspace{24mm} - 7 \, \nnu^2 \, m_u^2 \left( 9 \, m_u^2 + 5 \, m_W^2 - 5 \, m_Z^2 \right) + 19 \, \nnd^2 \, m_d^2 \left( m_W^2 - m_Z^2 \right) \Big] \bigg\}, \\[0.5em]
\dfrac{C_{\phi q}^{(3)}}{\Lambda^2} &=& \dfrac{G_F^2}{360 \, \pi^2 \, \Lambda^2} \bigg(5 \cot^2\beta \Big[ -3 \left( \nnd^2 \, m_d^4  - 6 \, \nnd \, \nnu \, \, m_d^2 \, m_u^2 + \nnu^2 \, m_u^4 \right) \nonumber \\
&& \hspace{24mm} + \left( \nnd^2 \,  m_d^2 + \nnu^2 \, m_u^2 \right) \, m_W^2 \Big] -3 m_W^4\bigg), \\[0.5em]
\dfrac{C_{\phi l}^{(1)}}{\Lambda^2} &=& \dfrac{G_F^2}{360 \, \pi^2 \, \Lambda^2} \Big[ 3 \left( m_W^2 - m_Z^2 \right) ^2 - 5 \, m_{e}^2 \, \nnl^2 \, \left( 21  \, m_{e}^2 + 17 \left(  m_W^2 - m_Z^2 \right) \right) \, \cot^2 \beta \Big], \\[0.5em]
\dfrac{C_{\phi l}^{(3)}}{\Lambda^2} &=& -\dfrac{G_F^2}{360 \, \pi^2 \, \Lambda^2} \Big[ 3 \, m_W^4 + 5 \, m_{e}^2 \nnl^2 \left( 3 \, m_{e}^2 - m_W^2 \right) \, \cot^2\beta \Big], \\[0.5em]
\dfrac{C_{\bm{ll}}}{\Lambda^2} &=& - \dfrac{m_W^4 \, G_F^2}{120 \, \pi^2 \, \Lambda^2}.
\eea
\es

\section{Higgs observables to \texorpdfstring{$\mathcal{O}(\xi^2)$}{O(xi^2)}}
\label{app:Higgs}

Here we summarize the dependence of the Higgs signal strengths on the various WCs generated in the 2HDM. We include the dependence on coefficients up to $\mathcal{O}(\xi^2)$, generated at tree-level, as well as those generated at one-loop at $\mathcal{O}(\xi^1)$. 
We emphasize that these expressions do not include the most general dependence on the SMEFT coefficients, regardless of flavor \textit{ansatz}, as we omit any operators that are not generated in the 2HDM, which are irrelevant for our analysis.

We consider the signal strengths for Higgs production in gluon-gluon fusion (ggF), vector boson-fusion (VBF), associated production with a $W$ or $Z$ boson ($Wh$ and $Zh$), and $t\bar{t}h$ production. We neglect any SMEFT deviations in the single-top ($th$) production mode, which is often measured together with $t\bar{t}h$ production, and assume that the $t\bar{t}h$ contribution is dominant. 
The signal strengths for Higgs production are defined as $\mu_{\textrm{prod}} = \sigma_{\textrm{prod}}^{\textsc{SMEFT}} / \sigma_{\textrm{prod}}^{\textsc{SM}}$, where $\sigma_{\textrm{prod}}^{\textsc{SM}}$ is the SM value.
When multiple production channels are combined into a single measurement, we take the signal strength to be the average of the two channels, weighted by their relative SM predictions from Ref.~\cite{LHCHiggsCrossSectionWorkingGroup:2013rie}. 
We indicate $8\,\textrm{TeV}$ or $13\,\textrm{TeV}$ for the signal strengths that depend on the collider energy.

The signal strengths of the decays shown are the same ratio with the partial widths evaluated in the SMEFT and in the SM. We consider Higgs decays to $b\bar{b}$, $WW^*(\to \ell\nu\ell\nu)$, $gg$, $\tau^+\tau^-$, $ZZ^*(\to 4\ell)$, $\gamma\gamma$, $Z\gamma$ and $\mu^+\mu^-$. 
For all the decays involving either on- or off-shell $Z$ and $W$'s, we neglect any effective operators appearing in the $Z, W \to f\bar{f}$ vertex, so that, e.g., $\mu_{h \to 4\ell} = \mu_{h \to \ell^+\ell^-\nu\bar{\nu}} = \mu_{h \to \ell^+\ell^- q\bar{q}} \equiv \mu_{h \to ZZ^*}$. 
The signal strengths for the individual decay modes can be combined with the known, SM branching ratios (BRs) for the Higgs to get the signal strength for the total width,
\begin{equation}
\mu_{\Gamma_{h,\textrm{tot.}}} = \sum_X \,
\mu_{h \to X} \times \textrm{BR}(h \to X)_{\textsc{SM}} ,
\end{equation}
which can then be used to predict the individual BRs in the SMEFT. We take the predictions for the SM branching ratios from Ref.~\cite{LHCHiggsCrossSectionWorkingGroup:2013rie}.

In computing the coefficients below, we use 
$G_F = 1.1663787 \times 10^{-5}\,\textrm{GeV}^{-2}$, $m_Z = 91.1876\,\textrm{GeV}$, $m_W = 80.379\,\textrm{GeV}$, $m_h = 125.0\,\textrm{GeV}$, $m_t = 173\,\textrm{GeV}$, $m_b = 4.18\,\textrm{GeV}$ and $m_{\tau} = 1.776\,\textrm{GeV}$. We keep only the masses of the third generation fermions, and set the first and second generation masses to zero. The WCs are given in units of $\textrm{TeV}^{-2}$ for the dimension-six operators, and $\textrm{TeV}^{-4}$ for those at dimension-eight. The signal strengths for $h\rightarrow 4\ell$ are taken from Ref.~\cite{ATLAS:2019dhi} to account for the effects of experimental efficiencies, while the others are computed analytically at leading order.

\subsection{Production}

\bs
\bea
\mu_{\textrm{ggF}} 
& = & 1 
+ 0.249 \, C_{d \phi,33} 
+ 0.121 \, C_{\phi\Box} 
- 0.129 \, C_{u \phi,33} 
- 0.0606 \, C_{\phi l,11}^{(3)} 
- 0.0606 \, C_{\phi l,22}^{(3)} 
\nonumber\\
&& \quad 
+\; 0.0606 \, C_{\bm{ll}} 
- 0.0303 \, C_{\phi D} 
+ 0.0540 \, (C_{d \phi,33})^2 
- 0.0182 \, C_{d \phi,33} \, C_{u\phi,33} 
\nonumber\\
&& \quad  
+\; 0.00421 \, (C_{u\phi,33})^2 
+ 0.0151 \, C_{qd\phi^5,33} 
- 0.00784 \, C_{qu\phi^5,33}, 
\\[0.5em]
\mu_{\textsc{vbf}}^{\textrm{8\,TeV}} 
& = & 1 
- 0.370 \, C_{\phi q,11}^{(3)} 
- 0.344 \, C_{\bm{\phi l}}^{(3)} 
+ 0.113 \, C_{\phi\Box} 
+ 0.0825 \, C_{\bm{ll}} 
\nonumber\\
&& \quad 
-\; 0.0533 \, C_{\phi W} 
+ 0.0238 \, C_{\phi WB} 
- 0.0148 \, C_{\phi u,11} 
+ 0.0112 \, C_{\phi q,11}^{(1)} 
\nonumber\\
&& \quad 
-\; 0.0106 \, C_{\phi D} 
+ 0.00353 \, C_{\phi d,11} 
- 0.00304 \, C_{\phi B} 
+ 0.000392 \, C_{\phi^6}^{(1)},
\\[0.5em]
\mu_{\textsc{vbf}}^{\textrm{13\,TeV}}
& = & 1 
- 0.423 \, C_{\phi q,11}^{(3)} 
- 0.347 \, C_{\phi q,11}^{(3)} 
+ 0.1005 \, C_{\phi\Box} 
+ 0.0826 \, C_{\bm{ll}} 
\nonumber\\
&& \quad 
-\; 0.0670 \, C_{\phi W} 
- 0.02955 \, C_{\phi u,11} 
- 0.0150 \, C_{\phi D} 
+ 0.0126 \, C_{\phi WB} 
\nonumber\\
&& \quad 
-\; 0.0107 \, C_{\phi B} 
+ 0.00893 \, C_{\phi q,11}^{(1)} 
+ 0.00313 \, C_{\phi d,11} 
+ 0.000490 \, C_{\phi^6}^{(1)},
\\[0.5em]
\mu_{Wh}^{\textrm{8\,TeV}} 
& = & 1 
+ 1.819 \, C_{\phi q,11}^{(3)} 
+ 0.8775 \, C_{\phi W} 
+ 0.1214 \, C_{\phi\Box} 
+ 0.0602 \, C_{\bm{ll}} 
\nonumber\\
&& \quad 
-\; 0.0309 \, C_{\phi D} 
- 7.916\times 10^{-6} \, C_{\phi^6}^{(1)},
\\[0.5em]
\mu_{Wh}^{\textrm{13\,TeV}}
& = & 1 
+ 1.950 \, C_{\phi q,11}^{(3)} 
+ 0.887 \, C_{\phi W} 
+ 0.1217 \, C_{\phi\Box} 
+ 0.0606 \, C_{\bm{ll}} 
\nonumber\\
&& \quad 
-\; 0.0303 \, C_{\phi D} 
- 4.326\times 10^{-6} \, C_{\phi^6}^{(1)},
\\[0.5em]
\mu_{Zh}^{\textrm{8\,TeV}}
& = & 1 
+ 1.716 \, C_{\phi q,11}^{(3)} 
+ 0.721 \, C_{\phi W} 
+ 0.426 \, C_{\phi u,11}
+ 0.314 \, C_{\phi WB} 
- 0.173 \, C_{\phi q,11}^{(1)} 
\nonumber\\
&& \quad 
-\; 0.142 \, C_{\phi d,11} 
+ 0.121 \, C_{\phi\Box} 
+ 0.0865 \, C_{\phi B} 
+ 0.06045 \, C_{\bm{ll}} 
+ 0.0375 \, C_{\phi D}
\nonumber\\
&& \quad 
-\; 3.515\times 10^{-6} \, C_{\phi^6}^{(1)},
\\[0.5em]
\mu_{Zh}^{\textrm{13\,TeV}} 
& = & 1 
+ 1.716 \,C_{\phi q,11}^{(3)} 
+ 0.721 \,C_{\phi W} 
+ 0.426 \,C_{\phi u,11} 
- 0.173 \,C_{\phi q,11}^{(1)} 
- 0.142 \,C_{\phi d,11} 
\nonumber\\
&& \quad 
+\; 0.121 \,C_{\phi\Box} 
+ 0.0865 \,C_{\phi B} 
+ 0.0375 \,C_{\phi D} 
+ 0.314 \,C_{\phi WB} 
+ 0.06045 \,C_{\bm{ll}} 
\nonumber\\
&& \quad 
-\; 3.515 \times 10^{-6} \, C_{\phi^6}^{(1)},
\\[0.5em]
\mu_{t\bar{t}h} 
& = & 1 
+ 0.121 \,C_{\phi\Box} 
- 0.122 \,C_{u\phi,33} 
- 0.0606 \,C_{\phi l,11}^{(3)} 
- 0.0606\,C_{\phi l,22}^{(3)} 
+ 0.0606 \,C_{\bm{ll}} 
\nonumber\\
&& \quad 
-\; 0.0303\,C_{\phi D} 
- 0.00740 \,C_{qu\phi^5,33} 
+ 0.00372 \,(C_{u\phi,33})^2. 
\eea
\es

\subsection{Decays}

\bs
\bea
\mu_{h \to b\bar{b}} & = & 1 
- 5.050 \, C_{d\phi,33} 
+ 0.121 \, {C_{\phi \Box}} 
- 0.121 \, {C_{\bm{\phi l}}^{(3)}}
- 0.0303 \, {C_{\phi D}} 
+ 0.0606 \, {C_{\bm{ll}}}, 
\nonumber\\
&& \quad 
+\; 6.376 \, (C_{d\phi,33})^2 
- 0.306 \, C_{qd\phi^5,33}, 
\\[0.5em]%
\mu_{h \to WW^*} & = & 1 + 
0.1202 \, {C_{\phi \Box}} 
+ 0.0935 \, {C_{\bm{ll}}} 
- 0.0895 \,C_{\phi W} 
- 0.0297 \,{C_{\phi D}} 
+ 0.000507 \, C_{\phi ^6}^{(1)},
\\[0.5em]%
\mu_{h \to gg} & = & 1 
+ 0.249 \, C_{d\phi,33} 
- 0.129 \, C_{u\phi,33} 
+ 0.1225 \, {C_{\phi \Box}} 
+ 0.0613 \, {C_{\bm{ll}}}  
- 0.0306 \, {C_{\phi D}} 
\nonumber\\
&& \quad
+\; 0.054 \, (C_{d\phi,33})^2 
- 0.0182 \, C_{d\phi,33} \, C_{u\phi,33} 
+ 0.0042 \, C_{u\phi,33}^2 
\nonumber\\
&& \quad 
-\; 0.0078 \, C_{qu\phi^5,33} 
+ 0.0151 \, C_{qd\phi^5,33},
\\[0.5em]%
\mu_{h \to \tau^+\tau^-} & = & 1 
- 11.88 \, {C_{e \phi,33}} 
+ 0.121 \, {C_{\phi \Box}} 
- 0.121 \, {C_{\bm{\phi l}}^{(3)}} 
+ 0.0606 \, {C_{\bm{ll}}} 
- 0.0303 \, {C_{\phi D}}
\nonumber\\
&& \quad 
+\; 35.29 \, (C_{e \phi,33})^2 
- 0.720 \, {C_{e\phi^5,33}}, 
\\[0.5em]%
\mu_{h \to ZZ^*} & = & 1 
+ 0.296 \, C_{\phi WB} 
- 0.296 \, C_{\phi W} 
- 0.234 \, C_{\bm{\phi l}}^{(3)} 
- 0.197 \, C_{\phi B} 
+ 0.181 \, C_{\bm{ll}}
\nonumber\\
&& \quad 
+\; 0.119 \, C_{\phi \Box} 
+ 0.126 \, C_{\phi l,11}^{(1)} 
- 0.101 \, C_{\phi l,11} 
+ 0.005 \, C_{\phi D}, 
\\[0.5em]%
\mu_{h \to \gamma\gamma} & = & 1 
- 40.15 \, C_{\phi B} 
+ 22.30 \, C_{\phi W B} 
- 13.08 \, C_{\phi W} 
- 0.364 \, C_{\phi l,11}^{(3)} 
- 0.242 \, C_{\phi D}
\nonumber\\
&& \quad 
+\; 0.182 \, C_{\bm{ll}} 
+ 0.121 \, C_{\phi  \Box} 
+ 0.0345 \, C_{u \phi,33}, 
\\[0.5em]%
\mu_{h \to Z\gamma} & = & 1 
- 15.47 \, C_{\phi B} 
+ 14.58 \, C_{\phi W} 
- 11.04 \, C_{\phi WB} 
- 0.182 \, C_{\bm{\phi l}}^{(3)} 
\nonumber\\
&& \quad 
+\; 0.1215 \, C_{\phi \Box} 
- 0.121 \, C_{\phi D} 
+ 0.0177 \, C_{\phi q, 11}^{(1)} 
- 0.0177 \, C_{\phi q, 11}^{(3)} 
+ 0.0177 \, C_{\phi u,11} 
\nonumber\\
&& \quad 
+\; 0.0909 \, C_{\bm{ll}} 
+ 0.00721 \, C_{u\phi,33} 
+ 0.00184 \, C_{\phi^6}^{(1)} 
+ 0.000437 \, C_{qu \phi^5, 33},
\\[0.5em]
\mu_{h \to \mu^+\mu^-} & = & 1 
- 199.79 \, C_{e \phi,22} 
+ 0.121 \, {C_{\phi \Box}}
- 0.121 \, {C_{\bm{\phi l}}^{(3)}} 
+ 0.0606 \, {C_{\bm{ll}}} 
- 0.0303 \, {C_{\phi D}} 
\nonumber\\
&& \quad 
+\; 9978.95 \, (C_{e \phi,22})^2 
- 12.11 \, C_{e \phi^5,22}. 
\eea
\es

\addcontentsline{toc}{section}{References}
{\small
\bibliographystyle{utphys}
\bibliography{refs}

\providecommand{\href}[2]{#2}\begingroup\raggedright\begin{thebibliography}{10}

\bibitem{Brivio:2017vri}
I.~Brivio and M.~Trott, ``{The Standard Model as an Effective Field Theory},''
  \href{http://dx.doi.org/10.1016/j.physrep.2018.11.002}{{\em Phys. Rept.}
  {\bfseries 793} (2019) 1--98},
  \href{http://arxiv.org/abs/1706.08945}{{\ttfamily arXiv:1706.08945
  [hep-ph]}}.

\bibitem{Isidori:2023pyp}
G.~Isidori, F.~Wilsch, and D.~Wyler, ``{The Standard Model effective field
  theory at work},'' \href{http://arxiv.org/abs/2303.16922}{{\ttfamily
  arXiv:2303.16922 [hep-ph]}}.

\bibitem{Henning:2014wua}
B.~Henning, X.~Lu, and H.~Murayama, ``{How to use the Standard Model effective
  field theory},'' \href{http://dx.doi.org/10.1007/JHEP01(2016)023}{{\em JHEP}
  {\bfseries 01} (2016) 023}, \href{http://arxiv.org/abs/1412.1837}{{\ttfamily
  arXiv:1412.1837 [hep-ph]}}.

\bibitem{Gorbahn:2015gxa}
M.~Gorbahn, J.~M. No, and V.~Sanz, ``{Benchmarks for Higgs Effective Theory:
  Extended Higgs Sectors},''
  \href{http://dx.doi.org/10.1007/JHEP10(2015)036}{{\em JHEP} {\bfseries 10}
  (2015) 036}, \href{http://arxiv.org/abs/1502.07352}{{\ttfamily
  arXiv:1502.07352 [hep-ph]}}.

\bibitem{Jiang:2018pbd}
M.~Jiang, N.~Craig, Y.-Y. Li, and D.~Sutherland, ``{Complete one-loop matching
  for a singlet scalar in the Standard Model EFT},''
  \href{http://dx.doi.org/10.1007/JHEP02(2019)031}{{\em JHEP} {\bfseries 02}
  (2019) 031}, \href{http://arxiv.org/abs/1811.08878}{{\ttfamily
  arXiv:1811.08878 [hep-ph]}}. [Erratum: JHEP 01, 135 (2021)].

\bibitem{Haisch:2020ahr}
U.~Haisch, M.~Ruhdorfer, E.~Salvioni, E.~Venturini, and A.~Weiler, ``{Singlet
  night in Feynman-ville: one-loop matching of a real scalar},''
  \href{http://dx.doi.org/10.1007/JHEP04(2020)164}{{\em JHEP} {\bfseries 04}
  (2020) 164}, \href{http://arxiv.org/abs/2003.05936}{{\ttfamily
  arXiv:2003.05936 [hep-ph]}}. [Erratum: JHEP 07, 066 (2020)].

\bibitem{Dawson:2020oco}
S.~Dawson, S.~Homiller, and S.~D. Lane, ``{Putting standard model EFT fits to
  work},'' \href{http://dx.doi.org/10.1103/PhysRevD.102.055012}{{\em Phys. Rev.
  D} {\bfseries 102} no.~5, (2020) 055012},
  \href{http://arxiv.org/abs/2007.01296}{{\ttfamily arXiv:2007.01296
  [hep-ph]}}.

\bibitem{DasBakshi:2020ejz}
S.~Das~Bakshi, J.~Chakrabortty, C.~Englert, M.~Spannowsky, and P.~Stylianou,
  ``{$CP$ violation at ATLAS in effective field theory},''
  \href{http://dx.doi.org/10.1103/PhysRevD.103.055008}{{\em Phys. Rev. D}
  {\bfseries 103} no.~5, (2021) 055008},
  \href{http://arxiv.org/abs/2009.13394}{{\ttfamily arXiv:2009.13394
  [hep-ph]}}.

\bibitem{Contino:2016jqw}
R.~Contino, A.~Falkowski, F.~Goertz, C.~Grojean, and F.~Riva, ``{On the
  Validity of the Effective Field Theory Approach to SM Precision Tests},''
  \href{http://dx.doi.org/10.1007/JHEP07(2016)144}{{\em JHEP} {\bfseries 07}
  (2016) 144}, \href{http://arxiv.org/abs/1604.06444}{{\ttfamily
  arXiv:1604.06444 [hep-ph]}}.

\bibitem{Belusca-Maito:2016cay}
H.~B\'elusca-Ma\"\i{}to, {\em {Search for new physics at the LHC using Higgs
  Effective Field Theory}}.
\newblock PhD thesis, Orsay, 3, 2016.

\bibitem{Ellis:2019zex}
J.~Ellis, S.-F. Ge, H.-J. He, and R.-Q. Xiao, ``{Probing the scale of new
  physics in the $ZZ\gamma$ coupling at $e^+e^-$ colliders},''
  \href{http://dx.doi.org/10.1088/1674-1137/44/6/063106}{{\em Chin. Phys. C}
  {\bfseries 44} no.~6, (2020) 063106},
  \href{http://arxiv.org/abs/1902.06631}{{\ttfamily arXiv:1902.06631
  [hep-ph]}}.

\bibitem{Li:2020gnx}
H.-L. Li, Z.~Ren, J.~Shu, M.-L. Xiao, J.-H. Yu, and Y.-H. Zheng, ``{Complete
  set of dimension-eight operators in the standard model effective field
  theory},'' \href{http://dx.doi.org/10.1103/PhysRevD.104.015026}{{\em Phys.
  Rev. D} {\bfseries 104} no.~1, (2021) 015026},
  \href{http://arxiv.org/abs/2005.00008}{{\ttfamily arXiv:2005.00008
  [hep-ph]}}.

\bibitem{Murphy:2020rsh}
C.~W. Murphy, ``{Dimension-8 operators in the Standard Model Eective Field
  Theory},'' \href{http://dx.doi.org/10.1007/JHEP10(2020)174}{{\em JHEP}
  {\bfseries 10} (2020) 174}, \href{http://arxiv.org/abs/2005.00059}{{\ttfamily
  arXiv:2005.00059 [hep-ph]}}.

\bibitem{Corbett:2021eux}
T.~Corbett, A.~Helset, A.~Martin, and M.~Trott, ``{EWPD in the SMEFT to
  dimension eight},'' \href{http://dx.doi.org/10.1007/JHEP06(2021)076}{{\em
  JHEP} {\bfseries 06} (2021) 076},
  \href{http://arxiv.org/abs/2102.02819}{{\ttfamily arXiv:2102.02819
  [hep-ph]}}.

\bibitem{Degrande:2023iob}
C.~Degrande and H.-L. Li, ``{Impact of dimension-8 SMEFT operators on diboson
  productions},'' \href{http://dx.doi.org/10.1007/JHEP06(2023)149}{{\em JHEP}
  {\bfseries 06} (2023) 149}, \href{http://arxiv.org/abs/2303.10493}{{\ttfamily
  arXiv:2303.10493 [hep-ph]}}.

\bibitem{Corbett:2023qtg}
T.~Corbett, J.~Desai, O.~J.~P. \'Eboli, M.~C. Gonzalez-Garcia, M.~Martines, and
  P.~Reimitz, ``{Impact of dimension-eight SMEFT operators in the electroweak
  precision observables and triple gauge couplings analysis in universal
  SMEFT},'' \href{http://dx.doi.org/10.1103/PhysRevD.107.115013}{{\em Phys.
  Rev. D} {\bfseries 107} no.~11, (2023) 115013},
  \href{http://arxiv.org/abs/2304.03305}{{\ttfamily arXiv:2304.03305
  [hep-ph]}}.

\bibitem{Dawson:2021xei}
S.~Dawson, S.~Homiller, and M.~Sullivan, ``{Impact of dimension-eight SMEFT
  contributions: A case study},''
  \href{http://dx.doi.org/10.1103/PhysRevD.104.115013}{{\em Phys. Rev. D}
  {\bfseries 104} no.~11, (2021) 115013},
  \href{http://arxiv.org/abs/2110.06929}{{\ttfamily arXiv:2110.06929
  [hep-ph]}}.

\bibitem{Dawson:2022cmu}
S.~Dawson, D.~Fontes, S.~Homiller, and M.~Sullivan, ``{Role of dimension-eight
  operators in an EFT for the 2HDM},''
  \href{http://dx.doi.org/10.1103/PhysRevD.106.055012}{{\em Phys. Rev. D}
  {\bfseries 106} no.~5, (2022) 055012},
  \href{http://arxiv.org/abs/2205.01561}{{\ttfamily arXiv:2205.01561
  [hep-ph]}}.

\bibitem{Banerjee:2022thk}
U.~Banerjee, J.~Chakrabortty, C.~Englert, S.~U. Rahaman, and M.~Spannowsky,
  ``{Integrating out heavy scalars with modified equations of motion: Matching
  computation of dimension-eight SMEFT coefficients},''
  \href{http://dx.doi.org/10.1103/PhysRevD.107.055007}{{\em Phys. Rev. D}
  {\bfseries 107} no.~5, (2023) 055007},
  \href{http://arxiv.org/abs/2210.14761}{{\ttfamily arXiv:2210.14761
  [hep-ph]}}.

\bibitem{Ellis:2023zim}
J.~Ellis, K.~Mimasu, and F.~Zampedri, ``{Dimension-8 SMEFT analysis of minimal
  scalar field extensions of the Standard Model},''
  \href{http://dx.doi.org/10.1007/JHEP10(2023)051}{{\em JHEP} {\bfseries 10}
  (2023) 051}, \href{http://arxiv.org/abs/2304.06663}{{\ttfamily
  arXiv:2304.06663 [hep-ph]}}.

\bibitem{Anisha:2020ggj}
Anisha, S.~Das~Bakshi, J.~Chakrabortty, and S.~K. Patra, ``{Connecting
  electroweak-scale observables to BSM physics through EFT and Bayesian
  statistics},'' \href{http://dx.doi.org/10.1103/PhysRevD.103.076007}{{\em
  Phys. Rev. D} {\bfseries 103} no.~7, (2021) 076007},
  \href{http://arxiv.org/abs/2010.04088}{{\ttfamily arXiv:2010.04088
  [hep-ph]}}.

\bibitem{Du:2022vso}
Y.~Du, X.-X. Li, and J.-H. Yu, ``{Neutrino seesaw models at one-loop matching:
  discrimination by effective operators},''
  \href{http://dx.doi.org/10.1007/JHEP09(2022)207}{{\em JHEP} {\bfseries 09}
  (2022) 207}, \href{http://arxiv.org/abs/2201.04646}{{\ttfamily
  arXiv:2201.04646 [hep-ph]}}.

\bibitem{Liao:2022cwh}
Y.~Liao and X.-D. Ma, ``{One-loop matching of scotogenic model onto standard
  model effective field theory up to dimension 7},''
  \href{http://dx.doi.org/10.1007/JHEP12(2022)053}{{\em JHEP} {\bfseries 12}
  (2022) 053}, \href{http://arxiv.org/abs/2210.04270}{{\ttfamily
  arXiv:2210.04270 [hep-ph]}}.

\bibitem{Li:2023ohq}
X.~Li and S.~Zhou, ``{One-loop Matching of the Type-III Seesaw Model onto the
  Standard Model Effective Field Theory},''
  \href{http://arxiv.org/abs/2309.14702}{{\ttfamily arXiv:2309.14702
  [hep-ph]}}.

\bibitem{Lee:1973iz}
T.~D. Lee, ``{A Theory of Spontaneous T Violation},''
\href{http://dx.doi.org/10.1103/PhysRevD.8.1226}{{\em Phys. Rev.} {\bfseries
  D8} (1973) 1226--1239}.
%%CITATION = PHRVA,D8,1226;%%.

\bibitem{Branco:2011iw}
G.~C. Branco, P.~M. Ferreira, L.~Lavoura, M.~N. Rebelo, M.~Sher, and J.~P.
  Silva, ``{Theory and phenomenology of two-Higgs-doublet models},''
  \href{http://dx.doi.org/10.1016/j.physrep.2012.02.002}{{\em Phys. Rept.}
  {\bfseries 516} (2012) 1--102},
\href{http://arxiv.org/abs/1106.0034}{{\ttfamily arXiv:1106.0034 [hep-ph]}}.
%%CITATION = ARXIV:1106.0034;%%.

\bibitem{Gunion:2002zf}
J.~F. Gunion and H.~E. Haber, ``{The CP conserving two Higgs doublet model: The
  Approach to the decoupling limit},''
  \href{http://dx.doi.org/10.1103/PhysRevD.67.075019}{{\em Phys. Rev. D}
  {\bfseries 67} (2003) 075019},
  \href{http://arxiv.org/abs/hep-ph/0207010}{{\ttfamily arXiv:hep-ph/0207010}}.

\bibitem{DeAngelis:2023bmd}
S.~De~Angelis and G.~Durieux, ``{EFT matching from analyticity and
  unitarity},'' \href{http://arxiv.org/abs/2308.00035}{{\ttfamily
  arXiv:2308.00035 [hep-ph]}}.

\bibitem{Li:2023edf}
X.~Li and S.~Zhou, ``{One-loop Matching and Running via On-shell Amplitudes},''
  \href{http://arxiv.org/abs/2309.10851}{{\ttfamily arXiv:2309.10851
  [hep-ph]}}.

\bibitem{Aebischer:2023nnv}
L.~Allwicher {\em et~al.}, ``{Computing tools for effective field theories:
  SMEFT-Tools 2022 Workshop Report, 14\textendash{}16th September 2022,
  Z\"urich},'' \href{http://dx.doi.org/10.1140/epjc/s10052-023-12323-y}{{\em
  Eur. Phys. J. C} {\bfseries 84} no.~2, (2024) 170},
  \href{http://arxiv.org/abs/2307.08745}{{\ttfamily arXiv:2307.08745
  [hep-ph]}}.

\bibitem{Criado:2017khh}
J.~C. Criado, ``{MatchingTools: a Python library for symbolic effective field
  theory calculations},''
  \href{http://dx.doi.org/10.1016/j.cpc.2018.02.016}{{\em Comput. Phys.
  Commun.} {\bfseries 227} (2018) 42--50},
  \href{http://arxiv.org/abs/1710.06445}{{\ttfamily arXiv:1710.06445
  [hep-ph]}}.

\bibitem{DasBakshi:2018vni}
S.~Das~Bakshi, J.~Chakrabortty, and S.~K. Patra, ``{CoDEx: Wilson coefficient
  calculator connecting SMEFT to UV theory},''
  \href{http://dx.doi.org/10.1140/epjc/s10052-018-6444-2}{{\em Eur. Phys. J. C}
  {\bfseries 79} no.~1, (2019) 21},
  \href{http://arxiv.org/abs/1808.04403}{{\ttfamily arXiv:1808.04403
  [hep-ph]}}.

\bibitem{Carmona:2021xtq}
A.~Carmona, A.~Lazopoulos, P.~Olgoso, and J.~Santiago, ``{Matchmakereft:
  automated tree-level and one-loop matching},''
  \href{http://dx.doi.org/10.21468/SciPostPhys.12.6.198}{{\em SciPost Phys.}
  {\bfseries 12} no.~6, (2022) 198},
  \href{http://arxiv.org/abs/2112.10787}{{\ttfamily arXiv:2112.10787
  [hep-ph]}}.

\bibitem{Fuentes-Martin:2022jrf}
J.~Fuentes-Mart\'\i{}n, M.~K\"onig, J.~Pag\`es, A.~E. Thomsen, and F.~Wilsch,
  ``{A proof of concept for matchete: an automated tool for matching effective
  theories},'' \href{http://dx.doi.org/10.1140/epjc/s10052-023-11726-1}{{\em
  Eur. Phys. J. C} {\bfseries 83} no.~7, (2023) 662},
  \href{http://arxiv.org/abs/2212.04510}{{\ttfamily arXiv:2212.04510
  [hep-ph]}}.

\bibitem{Gherardi:2020det}
V.~Gherardi, D.~Marzocca, and E.~Venturini, ``{Matching scalar leptoquarks to
  the SMEFT at one loop},''
  \href{http://dx.doi.org/10.1007/JHEP07(2020)225}{{\em JHEP} {\bfseries 07}
  (2020) 225}, \href{http://arxiv.org/abs/2003.12525}{{\ttfamily
  arXiv:2003.12525 [hep-ph]}}. [Erratum: JHEP 01, 006 (2021)].

\bibitem{Fuentes-Martin:2020udw}
J.~Fuentes-Martin, M.~K\"onig, J.~Pag\`es, A.~E. Thomsen, and F.~Wilsch,
  ``{SuperTracer: A Calculator of Functional Supertraces for One-Loop EFT
  Matching},'' \href{http://dx.doi.org/10.1007/JHEP04(2021)281}{{\em JHEP}
  {\bfseries 04} (2021) 281}, \href{http://arxiv.org/abs/2012.08506}{{\ttfamily
  arXiv:2012.08506 [hep-ph]}}.

\bibitem{Chen:2019pkq}
N.~Chen, T.~Han, S.~Li, S.~Su, W.~Su, and Y.~Wu, ``{Type-I 2HDM under the Higgs
  and Electroweak Precision Measurements},''
  \href{http://dx.doi.org/10.1007/JHEP08(2020)131}{{\em JHEP} {\bfseries 08}
  (2020) 131}, \href{http://arxiv.org/abs/1912.01431}{{\ttfamily
  arXiv:1912.01431 [hep-ph]}}.

\bibitem{Chen:2018shg}
N.~Chen, T.~Han, S.~Su, W.~Su, and Y.~Wu, ``{Type-II 2HDM under the Precision
  Measurements at the $Z$-pole and a Higgs Factory},''
  \href{http://dx.doi.org/10.1007/JHEP03(2019)023}{{\em JHEP} {\bfseries 03}
  (2019) 023}, \href{http://arxiv.org/abs/1808.02037}{{\ttfamily
  arXiv:1808.02037 [hep-ph]}}.

\bibitem{Dawson:2021jcl}
S.~Dawson, P.~P. Giardino, and S.~Homiller, ``{Uncovering the High Scale Higgs
  Singlet Model},'' \href{http://dx.doi.org/10.1103/PhysRevD.103.075016}{{\em
  Phys. Rev. D} {\bfseries 103} no.~7, (2021) 075016},
  \href{http://arxiv.org/abs/2102.02823}{{\ttfamily arXiv:2102.02823
  [hep-ph]}}.

\bibitem{auxiliary:2024bdfh}
S.~D. Bakshi, S.~Dawson, D.~Fontes, and S.~Homiller, ``Supplementary files for
  the 2hdm implementations.'' \url{https://github.com/BDFH-2024/BDFH}.

\bibitem{Gunion:1989we}
J.~F. Gunion, H.~E. Haber, G.~L. Kane, and S.~Dawson, ``{The Higgs Hunter's
  Guide},''
{\em Front. Phys.} {\bfseries 80} (2000) 1--404.
%%CITATION = FRPHA,80,1;%%.

\bibitem{Fontes:2021znm}
D.~Fontes, M.~L\"oschner, J.~C. Rom\~ao, and J.~P. Silva, ``{Leaks of CP
  violation in the real two-Higgs-doublet model},''
  \href{http://dx.doi.org/10.1140/epjc/s10052-021-09332-0}{{\em Eur. Phys. J.
  C} {\bfseries 81} no.~6, (2021) 541},
  \href{http://arxiv.org/abs/2103.05002}{{\ttfamily arXiv:2103.05002
  [hep-ph]}}.

\bibitem{Ginzburg:2005dt}
I.~F. Ginzburg and I.~P. Ivanov, ``{Tree-level unitarity constraints in the
  most general 2HDM},''
  \href{http://dx.doi.org/10.1103/PhysRevD.72.115010}{{\em Phys. Rev. D}
  {\bfseries 72} (2005) 115010},
  \href{http://arxiv.org/abs/hep-ph/0508020}{{\ttfamily arXiv:hep-ph/0508020}}.

\bibitem{Ivanov:2015nea}
I.~P. Ivanov and J.~P. Silva, ``{Tree-level metastability bounds for the most
  general two Higgs doublet model},''
  \href{http://dx.doi.org/10.1103/PhysRevD.92.055017}{{\em Phys. Rev.}
  {\bfseries D92} no.~5, (2015) 055017},
\href{http://arxiv.org/abs/1507.05100}{{\ttfamily arXiv:1507.05100 [hep-ph]}}.
%%CITATION = ARXIV:1507.05100;%%.

\bibitem{Haller:2018nnx}
J.~Haller, A.~Hoecker, R.~Kogler, K.~M\"onig, T.~Peiffer, and J.~Stelzer,
  ``{Update of the global electroweak fit and constraints on two-Higgs-doublet
  models},'' \href{http://dx.doi.org/10.1140/epjc/s10052-018-6131-3}{{\em Eur.
  Phys. J. C} {\bfseries 78} no.~8, (2018) 675},
  \href{http://arxiv.org/abs/1803.01853}{{\ttfamily arXiv:1803.01853
  [hep-ph]}}.

\bibitem{Brehmer:2015rna}
J.~Brehmer, A.~Freitas, D.~Lopez-Val, and T.~Plehn, ``{Pushing Higgs Effective
  Theory to its Limits},''
  \href{http://dx.doi.org/10.1103/PhysRevD.93.075014}{{\em Phys. Rev. D}
  {\bfseries 93} no.~7, (2016) 075014},
  \href{http://arxiv.org/abs/1510.03443}{{\ttfamily arXiv:1510.03443
  [hep-ph]}}.

\bibitem{Egana-Ugrinovic:2015vgy}
D.~Egana-Ugrinovic and S.~Thomas, ``{Effective Theory of Higgs Sector Vacuum
  States},'' \href{http://arxiv.org/abs/1512.00144}{{\ttfamily arXiv:1512.00144
  [hep-ph]}}.

\bibitem{Belusca-Maito:2016dqe}
H.~B\'elusca-Ma\"\i{}to, A.~Falkowski, D.~Fontes, J.~C. Rom\~ao, and J.~P.
  Silva, ``{Higgs EFT for 2HDM and beyond},''
  \href{http://dx.doi.org/10.1140/epjc/s10052-017-4745-5}{{\em Eur. Phys. J. C}
  {\bfseries 77} no.~3, (2017) 176},
  \href{http://arxiv.org/abs/1611.01112}{{\ttfamily arXiv:1611.01112
  [hep-ph]}}.

\bibitem{Belusca-Maito:2017iob}
H.~B\'elusca-Ma\"\i{}to, A.~Falkowski, D.~Fontes, J.~C. Rom\~ao, and J.~P.
  Silva, ``{CP violation in 2HDM and EFT: the $ZZZ$ vertex},''
  \href{http://dx.doi.org/10.1007/JHEP04(2018)002}{{\em JHEP} {\bfseries 04}
  (2018) 002}, \href{http://arxiv.org/abs/1710.05563}{{\ttfamily
  arXiv:1710.05563 [hep-ph]}}.

\bibitem{Banta:2023prj}
I.~Banta, T.~Cohen, N.~Craig, X.~Lu, and D.~Sutherland, ``{Effective field
  theory of the two Higgs doublet model},''
  \href{http://dx.doi.org/10.1007/JHEP06(2023)150}{{\em JHEP} {\bfseries 06}
  (2023) 150}, \href{http://arxiv.org/abs/2304.09884}{{\ttfamily
  arXiv:2304.09884 [hep-ph]}}.

\bibitem{Dawson:2023ebe}
S.~Dawson, D.~Fontes, C.~Quezada-Calonge, and J.~J. Sanz-Cillero, ``{Matching
  the 2HDM to the HEFT and the SMEFT: Decoupling and perturbativity},''
  \href{http://dx.doi.org/10.1103/PhysRevD.108.055034}{{\em Phys. Rev. D}
  {\bfseries 108} no.~5, (2023) 055034},
  \href{http://arxiv.org/abs/2305.07689}{{\ttfamily arXiv:2305.07689
  [hep-ph]}}.

\bibitem{Dawson:2023oce}
S.~Dawson, D.~Fontes, C.~Quezada-Calonge, and J.~J. Sanz-Cillero, ``{Is the
  HEFT matching unique?},'' \href{http://arxiv.org/abs/2311.16897}{{\ttfamily
  arXiv:2311.16897 [hep-ph]}}.

\bibitem{Arco:2023sac}
F.~Arco, D.~Domenech, M.~J. Herrero, and R.~A. Morales, ``{Nondecoupling
  effects from heavy Higgs bosons by matching 2HDM to HEFT amplitudes},''
  \href{http://dx.doi.org/10.1103/PhysRevD.108.095013}{{\em Phys. Rev. D}
  {\bfseries 108} no.~9, (2023) 095013},
  \href{http://arxiv.org/abs/2307.15693}{{\ttfamily arXiv:2307.15693
  [hep-ph]}}.

\bibitem{Buchalla:2023hqk}
G.~Buchalla, F.~K\"onig, C.~M\"uller-Salditt, and F.~Pandler, ``{Two-Higgs
  Doublet Model Matched to Nonlinear Effective Theory},''
  \href{http://arxiv.org/abs/2312.13885}{{\ttfamily arXiv:2312.13885
  [hep-ph]}}.

\bibitem{Haber:1989xc}
H.~E. Haber and Y.~Nir, ``{Multiscalar Models With a High-energy Scale},''
  \href{http://dx.doi.org/10.1016/0550-3213(90)90499-4}{{\em Nucl. Phys. B}
  {\bfseries 335} (1990) 363--394}.

\bibitem{Haber:2006ue}
H.~E. Haber and D.~O'Neil, ``{Basis-independent methods for the
  two-Higgs-doublet model. II. The Significance of tan$\beta$},''
  \href{http://dx.doi.org/10.1103/PhysRevD.74.015018}{{\em Phys. Rev. D}
  {\bfseries 74} (2006) 015018},
  \href{http://arxiv.org/abs/hep-ph/0602242}{{\ttfamily arXiv:hep-ph/0602242}}.
  [Erratum: Phys.Rev.D 74, 059905 (2006)].

\bibitem{Asner:2013psa}
D.~M. Asner {\em et~al.}, ``{ILC Higgs White Paper},'' in {\em {Community
  Summer Study 2013}: {Snowmass on the Mississippi}}.
\newblock 10, 2013.
\newblock \href{http://arxiv.org/abs/1310.0763}{{\ttfamily arXiv:1310.0763
  [hep-ph]}}.

\bibitem{Banerjee:2023iiv}
U.~Banerjee, J.~Chakrabortty, S.~U. Rahaman, and K.~Ramkumar, ``{One-loop
  effective action up to dimension eight: integrating out heavy scalar(s)},''
  \href{http://dx.doi.org/10.1140/epjp/s13360-024-04890-0}{{\em Eur. Phys. J.
  Plus} {\bfseries 139} no.~2, (2024) 159},
  \href{http://arxiv.org/abs/2306.09103}{{\ttfamily arXiv:2306.09103
  [hep-ph]}}.

\bibitem{Banerjee:2023xak}
U.~Banerjee, J.~Chakrabortty, S.~U. Rahaman, and K.~Ramkumar, ``{One-loop
  effective action up to any mass-dimension for non-degenerate scalars and
  fermions including light\textendash{}heavy mixing},''
  \href{http://dx.doi.org/10.1140/epjp/s13360-024-04966-x}{{\em Eur. Phys. J.
  Plus} {\bfseries 139} no.~2, (2024) 169},
  \href{http://arxiv.org/abs/2311.12757}{{\ttfamily arXiv:2311.12757
  [hep-ph]}}.

\bibitem{Grzadkowski:2010es}
B.~Grzadkowski, M.~Iskrzynski, M.~Misiak, and J.~Rosiek, ``{Dimension-Six Terms
  in the Standard Model Lagrangian},''
  \href{http://dx.doi.org/10.1007/JHEP10(2010)085}{{\em JHEP} {\bfseries 10}
  (2010) 085}, \href{http://arxiv.org/abs/1008.4884}{{\ttfamily arXiv:1008.4884
  [hep-ph]}}.

\bibitem{Jenkins:2013zja}
E.~E. Jenkins, A.~V. Manohar, and M.~Trott, ``{Renormalization Group Evolution
  of the Standard Model Dimension Six Operators I: Formalism and lambda
  Dependence},'' \href{http://dx.doi.org/10.1007/JHEP10(2013)087}{{\em JHEP}
  {\bfseries 10} (2013) 087}, \href{http://arxiv.org/abs/1308.2627}{{\ttfamily
  arXiv:1308.2627 [hep-ph]}}.

\bibitem{Jenkins:2013wua}
E.~E. Jenkins, A.~V. Manohar, and M.~Trott, ``{Renormalization Group Evolution
  of the Standard Model Dimension Six Operators II: Yukawa Dependence},''
  \href{http://dx.doi.org/10.1007/JHEP01(2014)035}{{\em JHEP} {\bfseries 01}
  (2014) 035}, \href{http://arxiv.org/abs/1310.4838}{{\ttfamily arXiv:1310.4838
  [hep-ph]}}.

\bibitem{Alonso:2013hga}
R.~Alonso, E.~E. Jenkins, A.~V. Manohar, and M.~Trott, ``{Renormalization Group
  Evolution of the Standard Model Dimension Six Operators III: Gauge Coupling
  Dependence and Phenomenology},''
  \href{http://dx.doi.org/10.1007/JHEP04(2014)159}{{\em JHEP} {\bfseries 04}
  (2014) 159}, \href{http://arxiv.org/abs/1312.2014}{{\ttfamily arXiv:1312.2014
  [hep-ph]}}.

\bibitem{Anisha:2021hgc}
Anisha, S.~Das~Bakshi, S.~Banerjee, A.~Biek\"otter, J.~Chakrabortty,
  S.~Kumar~Patra, and M.~Spannowsky, ``{Effective limits on single scalar
  extensions in the light of recent LHC data},''
  \href{http://dx.doi.org/10.1103/PhysRevD.107.055028}{{\em Phys. Rev. D}
  {\bfseries 107} no.~5, (2023) 055028},
  \href{http://arxiv.org/abs/2111.05876}{{\ttfamily arXiv:2111.05876
  [hep-ph]}}.

\bibitem{Dawson:2019clf}
S.~Dawson and P.~P. Giardino, ``{Electroweak and QCD corrections to $Z$ and $W$
  pole observables in the standard model EFT},''
  \href{http://dx.doi.org/10.1103/PhysRevD.101.013001}{{\em Phys. Rev. D}
  {\bfseries 101} no.~1, (2020) 013001},
  \href{http://arxiv.org/abs/1909.02000}{{\ttfamily arXiv:1909.02000
  [hep-ph]}}.

\bibitem{Asteriadis:2022ras}
K.~Asteriadis, S.~Dawson, and D.~Fontes, ``{Double insertions of SMEFT
  operators in gluon fusion Higgs boson production},''
  \href{http://dx.doi.org/10.1103/PhysRevD.107.055038}{{\em Phys. Rev. D}
  {\bfseries 107} no.~5, (2023) 055038},
  \href{http://arxiv.org/abs/2212.03258}{{\ttfamily arXiv:2212.03258
  [hep-ph]}}.

\bibitem{ParticleDataGroup:2022pth}
{\bfseries Particle Data Group} Collaboration, R.~L. Workman {\em et~al.},
  ``{Review of Particle Physics},''
  \href{http://dx.doi.org/10.1093/ptep/ptac097}{{\em PTEP} {\bfseries 2022}
  (2022) 083C01}.

\bibitem{Bellafronte:2023amz}
L.~Bellafronte, S.~Dawson, and P.~P. Giardino, ``{The importance of flavor in
  SMEFT Electroweak Precision Fits},''
  \href{http://dx.doi.org/10.1007/JHEP05(2023)208}{{\em JHEP} {\bfseries 05}
  (2023) 208}, \href{http://arxiv.org/abs/2304.00029}{{\ttfamily
  arXiv:2304.00029 [hep-ph]}}.

\bibitem{Peskin:1991sw}
M.~E. Peskin and T.~Takeuchi, ``{Estimation of oblique electroweak
  corrections},'' \href{http://dx.doi.org/10.1103/PhysRevD.46.381}{{\em Phys.
  Rev. D} {\bfseries 46} (1992) 381--409}.

\bibitem{ATLAS:2016neq}
{\bfseries ATLAS, CMS} Collaboration, G.~Aad {\em et~al.}, ``{Measurements of
  the Higgs boson production and decay rates and constraints on its couplings
  from a combined ATLAS and CMS analysis of the LHC pp collision data at $
  \sqrt{s}=7 $ and 8 TeV},''
  \href{http://dx.doi.org/10.1007/JHEP08(2016)045}{{\em JHEP} {\bfseries 08}
  (2016) 045}, \href{http://arxiv.org/abs/1606.02266}{{\ttfamily
  arXiv:1606.02266 [hep-ex]}}.

\bibitem{ATLAS:2022vkf}
{\bfseries ATLAS} Collaboration, G.~Aad {\em et~al.}, ``{A detailed map of
  Higgs boson interactions by the ATLAS experiment ten years after the
  discovery},'' \href{http://dx.doi.org/10.1038/s41586-022-04893-w}{{\em
  Nature} {\bfseries 607} no.~7917, (2022) 52--59},
  \href{http://arxiv.org/abs/2207.00092}{{\ttfamily arXiv:2207.00092
  [hep-ex]}}. [Erratum: Nature 612, E24 (2022)].

\bibitem{CMS:2022dwd}
{\bfseries CMS} Collaboration, A.~Tumasyan {\em et~al.}, ``{A portrait of the
  Higgs boson by the CMS experiment ten years after the discovery.},''
  \href{http://dx.doi.org/10.1038/s41586-022-04892-x}{{\em Nature} {\bfseries
  607} no.~7917, (2022) 60--68},
  \href{http://arxiv.org/abs/2207.00043}{{\ttfamily arXiv:2207.00043
  [hep-ex]}}.

\bibitem{Kanemura:2015mxa}
S.~Kanemura, M.~Kikuchi, and K.~Yagyu, ``{Fingerprinting the extended Higgs
  sector using one-loop corrected Higgs boson couplings and future precision
  measurements},''
  \href{http://dx.doi.org/10.1016/j.nuclphysb.2015.04.015}{{\em Nucl. Phys. B}
  {\bfseries 896} (2015) 80--137},
  \href{http://arxiv.org/abs/1502.07716}{{\ttfamily arXiv:1502.07716
  [hep-ph]}}.

\bibitem{Altenkamp:2017ldc}
L.~Altenkamp, S.~Dittmaier, and H.~Rzehak, ``{Renormalization schemes for the
  Two-Higgs-Doublet Model and applications to $h \rightarrow WW/ZZ \rightarrow
  4$ fermions},'' \href{http://dx.doi.org/10.1007/JHEP09(2017)134}{{\em JHEP}
  {\bfseries 09} (2017) 134}, \href{http://arxiv.org/abs/1704.02645}{{\ttfamily
  arXiv:1704.02645 [hep-ph]}}.

\bibitem{Kanemura:2017wtm}
S.~Kanemura, M.~Kikuchi, K.~Sakurai, and K.~Yagyu, ``{Gauge invariant one-loop
  corrections to Higgs boson couplings in non-minimal Higgs models},''
  \href{http://dx.doi.org/10.1103/PhysRevD.96.035014}{{\em Phys. Rev. D}
  {\bfseries 96} no.~3, (2017) 035014},
  \href{http://arxiv.org/abs/1705.05399}{{\ttfamily arXiv:1705.05399
  [hep-ph]}}.

\bibitem{Altenkamp:2017kxk}
L.~Altenkamp, S.~Dittmaier, and H.~Rzehak, ``{Precision calculations for $h \to
  WW/ZZ \to 4$ fermions in the Two-Higgs-Doublet Model with Prophecy4f},''
  \href{http://dx.doi.org/10.1007/JHEP03(2018)110}{{\em JHEP} {\bfseries 03}
  (2018) 110}, \href{http://arxiv.org/abs/1710.07598}{{\ttfamily
  arXiv:1710.07598 [hep-ph]}}.

\bibitem{ATLAS:2024lyh}
{\bfseries ATLAS} Collaboration, G.~Aad {\em et~al.}, ``{Interpretations of the
  ATLAS measurements of Higgs boson production and decay rates and differential
  cross-sections in $pp$ collisions at $\sqrt{s}=13$ TeV},''
  \href{http://arxiv.org/abs/2402.05742}{{\ttfamily arXiv:2402.05742
  [hep-ex]}}.

\bibitem{Han:2020lta}
T.~Han, S.~Li, S.~Su, W.~Su, and Y.~Wu, ``{Comparative Studies of 2HDMs under
  the Higgs Boson Precision Measurements},''
  \href{http://dx.doi.org/10.1007/JHEP01(2021)045}{{\em JHEP} {\bfseries 01}
  (2021) 045}, \href{http://arxiv.org/abs/2008.05492}{{\ttfamily
  arXiv:2008.05492 [hep-ph]}}.

\bibitem{Degrassi:2016wml}
G.~Degrassi, P.~P. Giardino, F.~Maltoni, and D.~Pagani, ``{Probing the Higgs
  self coupling via single Higgs production at the LHC},''
  \href{http://dx.doi.org/10.1007/JHEP12(2016)080}{{\em JHEP} {\bfseries 12}
  (2016) 080}, \href{http://arxiv.org/abs/1607.04251}{{\ttfamily
  arXiv:1607.04251 [hep-ph]}}.

\bibitem{Degrassi:2021uik}
G.~Degrassi, B.~Di~Micco, P.~P. Giardino, and E.~Rossi, ``{Higgs boson
  self-coupling constraints from single Higgs, double Higgs and Electroweak
  measurements},'' \href{http://dx.doi.org/10.1016/j.physletb.2021.136307}{{\em
  Phys. Lett. B} {\bfseries 817} (2021) 136307},
  \href{http://arxiv.org/abs/2102.07651}{{\ttfamily arXiv:2102.07651
  [hep-ph]}}.

\bibitem{LHCHiggsCrossSectionWorkingGroup:2013rie}
{\bfseries LHC Higgs Cross Section Working Group} Collaboration, J.~R. Andersen
  {\em et~al.}, ``{Handbook of LHC Higgs Cross Sections: 3. Higgs
  Properties},'' \href{http://arxiv.org/abs/1307.1347}{{\ttfamily
  arXiv:1307.1347 [hep-ph]}}.

\bibitem{ATLAS:2019dhi}
{\bfseries ATLAS} Collaboration, ``{Methodology for EFT interpretation of Higgs
  boson Simplified Template Cross-section results in ATLAS},'' 10, 2019.
\newblock \url{https://cds.cern.ch/record/2694284}.

\end{thebibliography}\endgroup

\printindex
}
\end{document}